\documentclass[10pt]{iopart}

\usepackage[utf8]{inputenc}

\usepackage{enumerate}

\usepackage{amssymb}
\usepackage{graphicx}

\usepackage{subfigure}

\usepackage{mathptmx}

\usepackage{url}

\def\phi{\varphi}
\def\eps{\varepsilon}
\def\be{\begin{equation}}
\def\ee{\end{equation}}

\def\bea{\begin{eqnarray}}
\def\eea{\end{eqnarray}}

\def\IR{\mathbb{R}}

\def\Ri{R_{\infty}}
\def\bM{\bar{M}}
\def\bT{\bar{T}}
\def\Ti{T_{\infty}}
\def\Tc{T_{\bar{c}}}
\def\Tvi{\bar{T}_{\infty}}
\def\Tvc{\bar{T}_{\bar{c}}}

\newcommand{\eqref}[1]{(\ref{#1})}

\begin{document}

\newtheorem{theorem}{Theorem}[section]
\newtheorem{notation}[theorem]{Notation}
\newtheorem{condition}[theorem]{Condition}
\newtheorem{definition}[theorem]{Definition}
\newtheorem{remark}[theorem]{Remark}
\newtheorem{lemma}[theorem]{Lemma}
\newtheorem{sublemma}[theorem]{Sublemma}
\newtheorem{proposition}[theorem]{Proposition}
\newtheorem{corollary}[theorem]{Corollary}
\newtheorem{conjecture}[theorem]{Conjecture}
\newtheorem{assumption}{Assumption}
\newtheorem{incom}{\textbf{!!!INNER COMMENT!!!}}

\renewcommand\theassumption{A\arabic{assumption}}

\title{Chaos and stability in a two-parameter family of convex billiard tables}

  \author{ Péter Bálint$^1$, Miklós Halász$^1$, Jorge Hernández-Tahuilán$^2$ and David P.~Sanders$^2$ }

\address{$^1$  Institute of Mathematics, Budapest University of Technology and Economics, Egry József u. 1, H-1111, Budapest, Hungary }
\eads{\mailto{pet@math.bme.hu} , \mailto{miklos.halasz@yahoo.co.uk}}

\address{$^2$ Departamento de Física, Facultad de Ciencias, Universidad Nacional Autónoma de México, Ciudad Universitaria, 04510 México D.F., Mexico}
\eads{\mailto{tahuilan@ciencias.unam.mx} , \mailto{dps@fciencias.unam.mx} }

\date{\today}

\begin{abstract}
We study, by numerical simulations and semi-rigorous arguments, a two-parameter family of
convex, two-dimensional billiard tables, generalizing the one-parameter class of oval billiards of Benettin--Strelcyn \cite{BS}.
We observe interesting dynamical phenomena when the billiard tables are continuously deformed
from the integrable circular billiard to different versions of completely-chaotic stadia. In particular, we conjecture that
a new class of ergodic billiard tables is obtained in certain regions of the two-dimensional parameter space, when the billiards are close to
skewed stadia. We provide heuristic arguments supporting this conjecture, and give numerical confirmation using the powerful method of
Lyapunov-weighted dynamics.

\end{abstract}

%  MSC CODEs: Primary: 37D50; Secondary: 37A25, 37J40, 37M25 %/ams{}

% \maketitle

\section{Introduction}

Billiard models are a class of Hamiltonian dynamical systems which  exhibit the full range of behaviour from completely integrable
to completely chaotic dynamics \cite{CM}. They consist of a point particle which collides elastically with the walls of a bounded region, the \emph{billiard table};
the shape of the table determines the type of dynamics which is observed.

Several classes of billiard tables have been studied which interpolate between completely integrable and completely chaotic dynamics,
including both one-parameter \cite{BS,HW,mushroom,Ma,Fo} and two-parameter \cite{dullin_two-parameter_1996} families.
These allow us to observe the transitions by which the typical phase space, which is a mixture of
ergodic, chaotic components and regular KAM islands, evolves from one extreme behaviour to the other.

% Several one-parameter classes of billiard tables have been studied which interpolate between completely integrable and completely chaotic dynamics \cite{BS,HW,mushroom,Ma,Fo}.
% This allows us to observe the transitions by which the typical phase space, which is a mixture of
% ergodic, chaotic components and regular KAM islands, evolves from one extreme behaviour to the other.

In the context of two-dimensional billiards, a popular starting point of such investigations is the Bunimovich stadium, which consists of two
semicircular arcs connected by two parallel segments. The case when these two segments are non-parallel (and, correspondingly, one of the arcs
is shorter, with the other being longer than a semicircle) is often referred to as the skewed stadium or squash billiard table. Stadia are known to be
ergodic and hyperbolic; however, as a consequence of so-called ``quasi-integrable'' phenomena, the hyperbolicity is non-uniform, and the dynamical
behaviour is very sensitive to perturbations of the boundary. There is an abundance of literature on stadium billiards;  some
works that are relevant to our discussion are refs.~\cite{B, M, CZ2, BG, De}; see also section~\ref{s:introbill} for a more detailed
description of stadia.

In this paper, we study a two-parameter set of two-dimensional billiard tables which generalizes the one-parameter family of
oval billiards studied in \cite{BS,HW} in a particular way; see section~\ref{s:squashes} for an explicit description.  Our models constitute a subcase of a rather general class of billiards introduced, but not studied in detail, by Hayli and Dumont in \cite{hayli_experiences_1986}. Our class is of particular interest since it includes as limiting cases an entire family of systems which are known to be ergodic, the skewed stadium billiards. Here these are generalised by deforming the sides of the stadia to circular arcs.

As in previous works on billiards formed from piecewise smooth components \cite{BS, HW, hayli_experiences_1986, dullin_two-parameter_1996, LopacLemonPRE2001, LopacParabolicOval2002, LopacEllipticalStadium2006, LopacTruncatedElliptical2010}, for a large set of parameter values, we find coexistence  of stability islands
and chaotic components in phase space. However,
%
%  (in the complement of the above-described open set of parameters), coexistence of stability islands
% and chaotic components may be observed in the phase space.
%
% % By introducing an additional parameter which scales the skewness of the perturbed stadium, we observe
% new dynamical phenomena.
we also find numerically that billiards which are sufficiently close to the limiting skewed stadia appear to have
% In particular, numerical simulations indicate that for an open set of parameter values there are
\emph{no} remaining stability islands  -- the phase
space is filled by a single large chaotic, ergodic component. This motivates our Conjecture~\ref{conj} concerning a new class of ergodic billiard tables.
Similar conjectures have been suggested for other billiard models of similar type -- see, e.g., \cite{dullin_two-parameter_1996}.

Our case is, however, different from previous studies in several respects.
% \begin{itemize}
Firstly, the new class introduced in this paper and  conjectured (for an open set of parameter values) to be ergodic   consists of \emph{convex} planar billiards. The issue of ergodicity versus KAM islands in convex billiards has been in the focus of continued interest for several decades. On the one hand, by Lazutkin's observation in \cite{L}, and its strengthening according to Douady \cite{douadyphd}, a convex planar billiard with at least $C^6$ boundary cannot be ergodic.  Furthermore, recent results by Bunimovich and Grigo \cite{Grigophd, Bunimovich_Grigo} show that elliptic islands arise in $C^2$ stadium-like billiards (billiard tables constructed from stadia by replacing the discontinuity of the curvature with a $C^2$ smoothening).
On the other hand, several classes of convex planar billiards (with some discontinuity points of the curvature) are proved to be ergodic (see section~\ref{s:introbill} for a partial list of references). A common feature of these examples is the defocusing mechanism, which requires that whenever a narrow beam of (initially) parallel rays completes a series of consecutive reflections on one of the smooth focusing components of the billiard boundary, it must pass through a conjugate point and become divergent before the next collision with the curved (non-flat) part of the boundary; see \cite{CM, Grigophd} for a more detailed description.

It is easy to see that in our examples defocusing cannot take place in this sense. Hence the mechanism that produces the (conjectured) ergodic behaviour must be different. We are aware of two other examples of ergodic planar billiards with focusing boundary components where defocusing is violated \cite{Bussolari_Lenci, Bunimovich_delMagno}; however, non-convexity of the billiard domain plays an important role in both cases. Similarly, numerical studies in \cite{dullin_two-parameter_1996, DullinBackerErgodicityLimaconNonlin2001} suggest ergodicity only for certain \emph{non}-convex domains.

Secondly, even though a rigorous proof is currently not available, we give, in addition to the simulated phase portraits, further evidence which strongly supports our conjecture. Heuristic arguments are provided in section~\ref{s:heur}, which rely on the similarity of the dynamics with those of the skewed stadia, and on the explicit analysis of sliding trajectories. Furthermore, the absence of islands is also tested numerically by the powerful method of Lyapunov-weighted dynamics in section~\ref{sec:lyap}.

The rest of this paper is organized as follows. In section~\ref{sec:mod}, we define the two-parameter family of generalised squashes and highlight its differences from related models. Section~\ref{sec:param-dependence} reports numerical results concerning the dependence of the dynamics on the two geometrical parameters, and presents the main conjecture on the existence of a new class of ergodic billiards. Section~\ref{sec:evidence-supporting-conjecture} presents heuristic arguments and further numerical evidence supporting the conjecture; and in section~\ref{sec:conclusions} we draw some conclusions.

\section{The model}\label{sec:mod}
We begin by defining the class of generalised squash billiard models.

\subsection{Convex billiard tables}\label{s:introbill}

Consider a convex, compact domain $Q\subset \IR^2$ bounded by a closed, piecewise-smooth curve $\Gamma=\partial Q$.
The motion of a point particle
that travels along straight lines with constant speed in the interior of $Q$, and bounces off elastically
(angle of reflection equals angle of incidence) when reaching the boundary $\Gamma$, is referred to as \emph{billiard dynamics}.

We investigate these dynamics in discrete time, that is, from collision to collision. The phase space is then
the cylinder $M=\Gamma\times [0,\pi]$, with $M\ni x=(k,\phi)$, where the configurational coordinate is the arc length $k$ along the boundary,
which satisfies $0\le k <|\Gamma|$ and describes the point of the closed curve $\Gamma$ at which the collision takes place, while the velocity
coordinate $0\le \phi \le \pi$ describes the angle that the outgoing velocity makes with the (positively oriented) tangent
line to $\Gamma$ at the point $k$.

Given $x\in M$, the position and velocity at the
next collision are uniquely determined, so that the billiard map $T:M\to M$ is well-defined. It is usual to visualize
$M$ as a rectangular domain in the plane, and the consecutive points along a trajectory of $T$ as
points in this domain. $T$ has a natural invariant measure
$\mu$, which is absolutely continuous with respect to Lebesgue measure on $M$, given by
$$d\mu=\rm{\,const\,} .  \sin\phi\, dk\,d\phi.$$
For further material on billiards in general we refer to the monographs  \cite{CM,T}.

The billiard map may show a surprisingly wide variety of dynamical phenomena for different choices
of the billiard table $\Gamma$. The best known case
is the billiard in a circle: for this geometry the dynamics are \emph{integrable}:
the angle of incidence $\phi$ is an integral of motion, the values of which label the
invariant curves \cite{CM}.

If $\Gamma$ is a $C^{553}$-smooth closed curve, then for
trajectories in the vicinity of the boundary the dynamics resemble, to some extent,
those of the circular billiard: Lazutkin \cite{L} showed that a positive measure set of the
phase space in a neighborhood of the boundaries $\phi=0$ and $\phi=\pi$ is foliated by invariant curves.
Later, Douady (\cite{douadyphd}) lowered the requirement for Lazutkin's result to $C^6$-smooth boundaries.

However, if $\Gamma$ has less smoothness, then the billiard can be completely chaotic and ergodic.
The first examples of such billiard tables are the celebrated stadia:
\begin{itemize}
\item the \emph{straight stadium} is formed by two identical semicircles, joined at
their endpoints by two parallel lines  along their common tangents;
\item the \emph{skewed stadium}, or squash table, is formed by two circular arcs of different radii $r<R$,
joined at their endpoints by non-parallel straight lines along their common tangents.
\end{itemize}

Stadia were introduced and their chaotic character first studied by Bunimovich in his famous paper \cite{B}. It is known that for these tables the billiard map $T$ is completely hyperbolic, i.e.\ there is one strictly positive and one strictly negative Lyapunov exponent for $\mu$-a.e.
$x\in M$, and $T$ is ergodic with respect to $\mu$; see \cite{CZ2, CM, CZ1} for a detailed description of stadia.

Stadia have three important characteristic features:
\begin{itemize}
 \item The table boundary $\Gamma$ is only \emph{piecewise} smooth: at the intersection
	points of the circular arcs and straight lines, the curvature of the boundary (the second derivative) is discontinuous. The
	resulting singularities play
	a crucial role in the dynamics of the billiard map, both for the stadia and for the tables
	investigated below.
  \item The map $T$ has many periodic points, all of which are hyperbolic.
	In the case of the circular billiard, all periodic points are parabolic. We will see below that in the generalised
	squash tables, the third class, elliptic periodic points, may also arise, typically giving rise to KAM
	islands in their vicinity.
  \item The billiard map $T$ in stadia is only \emph{non-uniformly}
	hyperbolic. The reason for this is the presence of so-called quasi-integrable --
	sliding, bouncing and diametrical -- trajectory segments of arbitrary length. When the geometry is perturbed,
	these quasi-integrable phenomena may (or may not) create islands of integrability; see below for further discussion.
\end{itemize}

The  mechanism that is responsible for the chaotic behaviour in stadium-shaped tables, known as the \emph{defocusing mechanism}, has also been observed and studied in other classes of billiard tables in works of Bunimovich, Donnay, Markarian, Sz\'asz
and Wojtkowski \cite{W,M1,D,Bu2,Sz,M2}.
However, the geometry of the billiards studied in this paper
is quite different from the geometry of any of these classes.

\subsection{The two-parameter family of generalised squashes}\label{s:squashes}

The family of convex billiard tables studied in this paper is described by two parameters, $b$ and $c$.
The table is built on a trapezoid; on each side of the trapezoid is placed a circular arc
joining its end-points, and adjacent arcs are constrained to meet with common tangents.
% A circular arc
% built from four circular arcs of different radii, the
% adjacent arcs meet with common tangent lines in the edges of a circular trapezoid.
The parameter $0\le b\le 1$ specifies the
geometry of the trapezoid, while $1\le c \le \infty$ specifies the ratio of the radii of the arcs.

\begin{figure}[tp]
\centering
\subfigure[Trapezoids for $b=0$, $0.2$, $0.4$, $0.6$ and $1.0$, from inside to outside.]{\includegraphics[scale=.2]{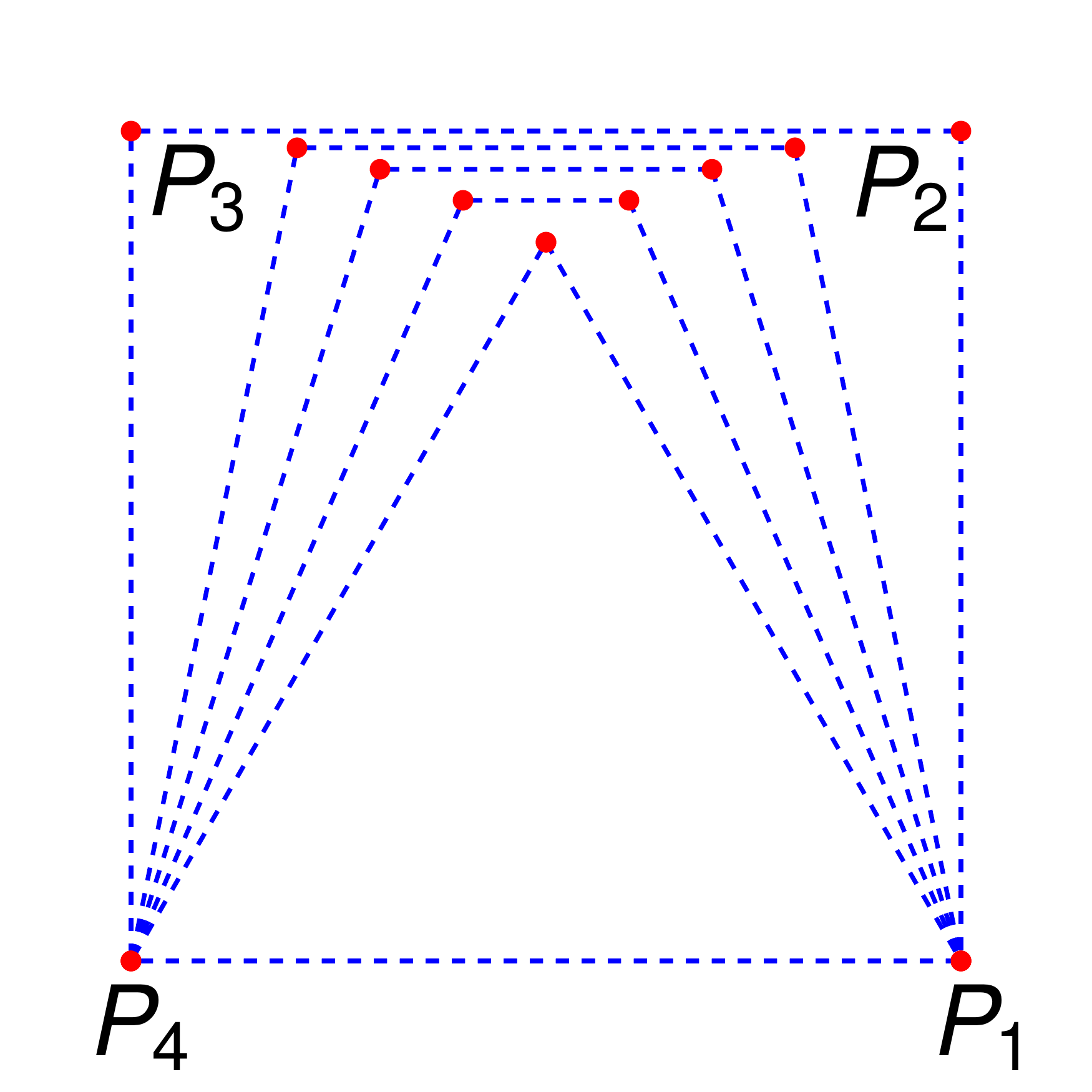}}
\quad
\subfigure[Effect of the parameter $c$, for $b=0.4$.]{\includegraphics[scale=.27 ]{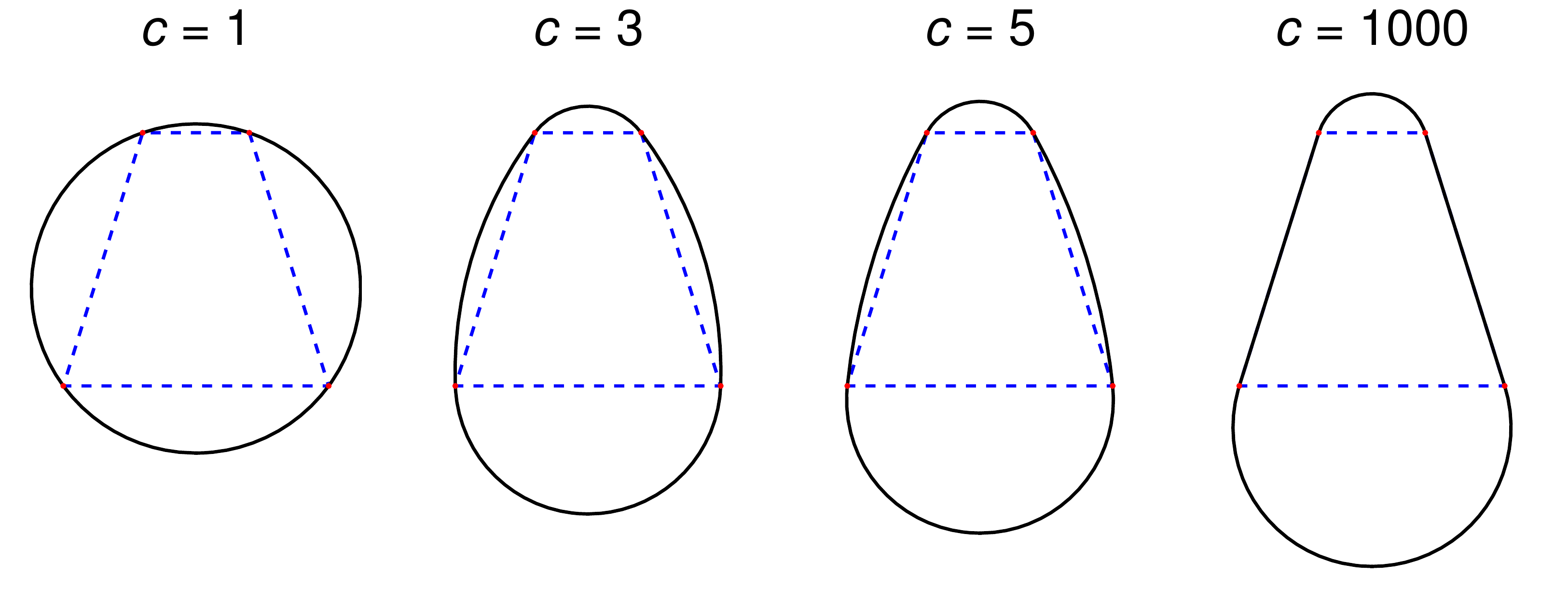}}
\caption{Variation of the geometry of the generalised squash billiards as a function of the two parameters $b$ and $c$.
}
\label{f:bc}
\end{figure}

More precisely, the base and the two sides of the trapezoid  are fixed to have unit length, and $b$ is the length of the top,
which is parallel to the base. The two extreme cases $b=0$ and $b=1$ correspond to
the equilateral triangle and the square, respectively; see figure~\ref{f:bc}(a).

The table is left--right symmetric, so that the circular arcs corresponding to the two sides have the same radius.
The radii of the arcs of the base, the sides and the top are denoted $R$, $R_{\infty}$ and $r$,
respectively.
%Note that the left and the right arcs have the same radii, thus the table has a left-right
%symmetry.
As adjacent arcs are required to have a common tangent line where they meet, %(in the relevant edge),
the shape of the table is determined by the value of any one of the radii -- i.e., for a given trapezoid (value of $b$),
the geometry can be parametrized by a single quantity. For convenience, we choose this parameter as the
ratio $c := R_{\infty}/R$; see figure~\ref{f:bc}. For brevity, throughout the paper the arcs corresponding to the base, the sides and the top
are referred to as the bottom arc, the ``almost-flat'' arcs, and the top arc, respectively.  The construction of the tables is detailed in the Appendix; they can be viewed as a subfamily of a general class of billiards introduced in \cite{hayli_experiences_1986}.

Note that the case $c=1$ corresponds to a circle for any value of $b$, the
case $b=1$ and $c=\infty$ corresponds to the straight stadium, while  $b < 1$ and $c=\infty$ gives a skewed
stadium, with the amount of skewness depending on the value of $b$. Thus, by changing $c$, we may
continuously deform the integrable dynamics of the circular billiard into a completely-chaotic billiard.

In this paper we concentrate on the role of the parameter $b$ in this transition, which determines the relative width of the top and bottom sections of the billiard.
The transition for the case $b=1$ (oval billiards based on a square) has already been studied in the literature, although with a different parametrisation
\cite{BS,HW}, where conclusions were drawn based on the phase portraits obtained.
% There, the authors
% performed simulations in fixed geometries, and drew conclusions
% based on the phase portraits obtained.
The picture for $b=1$ can be summarized as follows. As soon as
$c$ is finite  -- that is, the table differs from the straight stadium -- ergodicity of the billiard map
is ruined. More precisely, elliptic islands -- regions of positive Lebesgue %(or equivalently $\mu$)
measure, foliated by  invariant curves and concentrated around elliptic periodic points -- appear
in the phase space. For intermediate values $1<c<\infty$, coexistence of such elliptic islands
and ergodic components of positive measure is observed. For large enough $c$, there is just
a single such chaotic component, which still coexists with elliptic islands, while the number of chaotic components increases once
$c$ is decreased below certain threshholds. The references \cite{BS,HW} concentrated on this phenomenon of splitting of chaotic components.

In the present paper our main goal is different: by extending our investigation to $b<1$, we observe
distinct phenomena for finite, but large enough, $c$.
%Let us emphasize, for this reason, that

\subsection{Quasi-integrable phenomena}

\begin{figure}[tp]
\centering
\subfigure[Bouncing and sliding orbits.]{\includegraphics[scale=0.35]{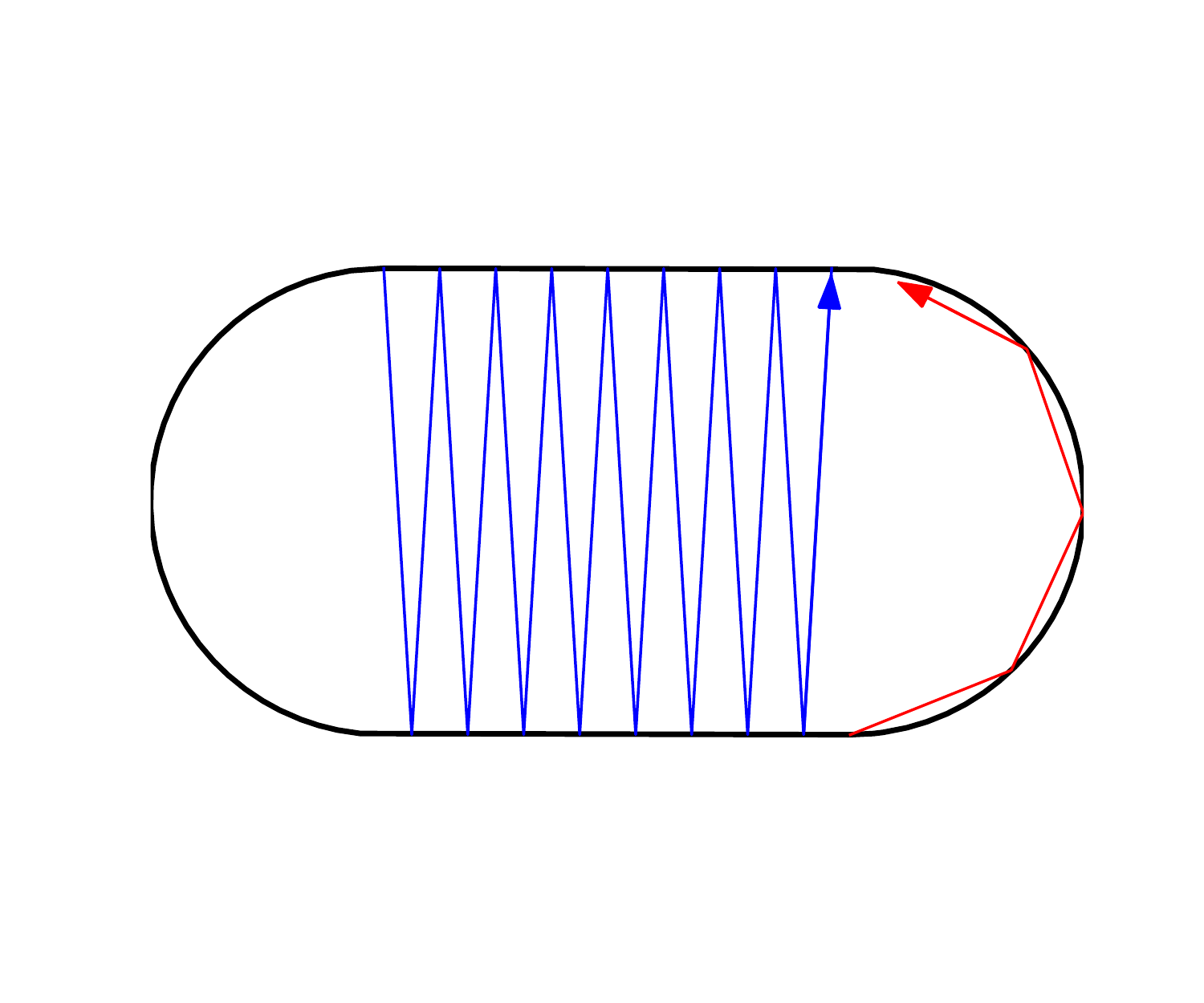}}\qquad\qquad
\subfigure[Diametrical orbits.]{\includegraphics[scale=0.4]{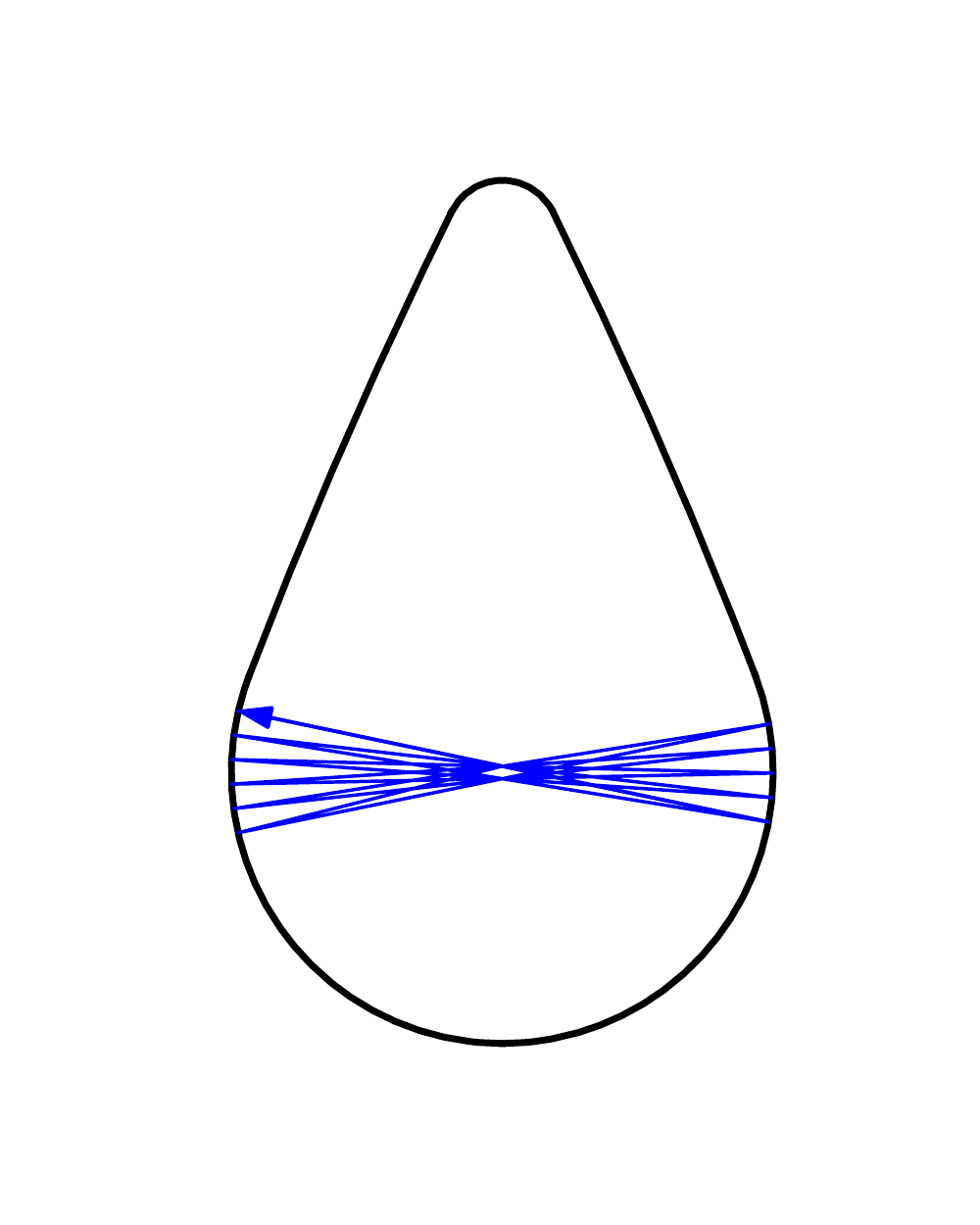}}

\caption{Different types of quasi-integrable phenomena present in stadia.}
\label{f:quasi}
\end{figure}

Certain differences between the $b=1$ and $b<1$ cases involve quasi-integrable trajectories, which, as mentioned above,
make stadia only non-uniformly hyperbolic. These
quasi-integrable phenomena have different geometric origins for straight and skewed stadia. Sliding -- where the
trajectory collides almost tangentially at a circular arc for many consecutive occasions -- is present in both
of these tables; however, bouncing -- where the trajectory travels back and forth between two flat boundary components
-- is characteristic only of the straight stadium; cf.\ figure~\ref{f:quasi}(a). More precisely,
the time of bouncing is uniformly bounded for skewed stadia, whereas it can last arbitrarily long in the
straight case. When $c<\infty$, the geometry is perturbed qualitatively at the almost-flat
sides, which is exactly where bouncing takes place. This makes the quasi-integrable
bouncing phenomena integrable -- an elliptic island is created -- if $b=1$. However, when $b<1$, the role of
bouncing is negligible, and so the changes are not so pronounced.

Finally, in skewed stadia a third type of quasi-integrable motion may also arise, in addition
to bouncing and sliding. When a circular arc is longer than a semi-circle, trajectories may travel back and forth
between two opposite points of the circular arc, with velocity almost parallel to the diameter,
for an arbitrarily long time; cf.\ figure~\ref{f:quasi}(b). However, for finite $c$, the qualitative changes in the geometry do
\emph{not} affect this diametrical motion, which thus remains a source of quasi-integrable phenomena, and does not create
islands of stability. See \cite{M,CZ2, BG,CZ} for discussions of the effect of quasi-integrable phenomena on the ergodic properties
of stadia.

\section{Parameter-dependence of dynamics}
\label{sec:param-dependence}

In this section, we survey the types of dynamical behaviour observed in different regions of parameter space. The numerical methods that were used are briefly described in the Appendix.

\subsection{Parameter space}
  \label{s:param_space}

We now proceed to investigate the dependence of the dynamics on the parameters of the system, by systematically scanning the two-dimensional parameter space and identifying the different phenomena observed. Numerical results for other billiard models of similar types, formed by piecewise-smooth curves, can be found in \cite{BS, HW, hayli_experiences_1986, dullin_two-parameter_1996, LopacLemonPRE2001, LopacParabolicOval2002, LopacEllipticalStadium2006, LopacTruncatedElliptical2010}.

For not too small $c$ (for $c \ge 1.2$ in all cases), ergodic components of positive Lebesgue measure seem to fill most of the phase space. An initial condition chosen randomly in such a component has an orbit which fills the component densely. For $c\ge 1.5$, we observe a single, dominant ergodic component, whereas for smaller values of $c$ several ergodic components coexist. For $b=1$ this ``splitting of chaotic components'', studied in \cite{BS,HW}, occurs at $c \simeq 1.35$; for $b<1$ we observe similar phenomena, although restricted to a smaller range of $c$.

If the initial point is chosen in the complement of the large ergodic component(s), then we observe the typical mixed phase space characteristic of KAM theory: regions centered around elliptic periodic points where a positive measure set foliated by invariant curves and chaotic orbits coexist in a topologically complicated way. We refer to these regions as ``islands''.

\subsection{Stability regions of periodic orbits}
% Period-$2$ orbits: comparison between $b=1$ and $b<1$}
To gain understanding of the observed phenomena, it is useful to start from
the known behaviour of the case $b=1$, i.e.\ the oval billiards studied in \cite{BS, HW}.
%, which are based on a square.

\begin{figure}[tp]
\centering
\includegraphics[scale=.54]{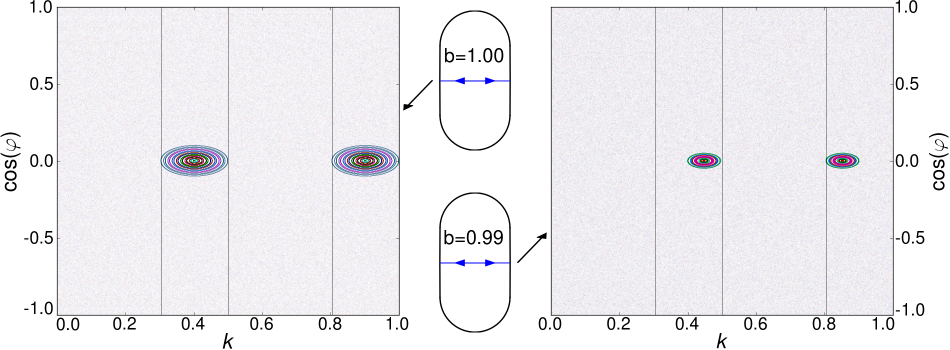}
\caption{The period-2 orbit and corresponding phase-space structures for c=100 with $b=1$ (left) and $b=0.99$ (right).}
\label{f:2per_stab}
\end{figure}

For these billiards, an important feature of the phase portrait is a period-$2$ orbit which is present for \emph{any} finite value of $c$, and which bounces
perpendicularly between the midpoints of the almost-flat sides. Its stability properties
can easily be calculated
% (cf.\ section~\ref{s:period_doubling_geom_destruction})
(cf.\ Appendix A): when $c<\infty$ (that is, $R_{\infty}<\infty$),
it is elliptic, and gives rise to a KAM island. The island around this orbit remains fully extended in the $k$ coordinate along the almost-flat arcs, and it shrinks only in the $\phi$ direction as we increase the value of $c$; see figure~\ref{f:2per_stab}(a).

For $b<1$, however, the island shrinks both in the $k$ and in the $\phi$ directions as $c$ is increased (see figure~\ref{f:2per_stab}(b)), and finally disappears for large enough $c$. The reason for this is that the almost-flat arcs are now placed along two \emph{non-parallel} sides of the trapezoid, so that for large enough $c$ there no longer exist any points on these two arcs which have mutually parallel tangent lines, and hence there is no period-$2$ orbit.
In this case, the bottom arc is forced to be longer than a semi-circle, due to the orientation of the tangent lines at the points where the bottom and almost-flat arcs join. The limit of existence of this period-$2$ orbit is thus when the bottom arc is exactly a semi-circle, which occurs at $c = c_0(b) := \frac{2}{1-b}$.
%  if the bottom arc
% is sufficiently curved, then it becomes longer than a semi-circle, which implies that there no longer exist points on the
% two almost-flat arcs with parallel tangent lines, in contrast to the $b=1$ case.
% % , where the tangents at
% % the midpoints of the two almost-flat arcs are parallel for any value of $c$.

% We point out the differences between the cases $b=1$ and $b<1$. For $b=1$, there is a large ergodic component with a single island in it, formulated around the elliptic period two orbit bouncing between the two almost flat arcs, cf.~section~\ref{sec:mod} and figure~\ref{f:2per_stab} below.

% For $b<1$, }. For large enough $c$ -- actually exactly when the lower arc becomes larger than a semi-circle -- it disappears completely.

% Hereby, whether the bottom arc is smaller or larger than a semi-circle plays an important role. In the region of the parameter space where the bottom arc is smaller than a semi-circle there is an island around a period--two orbit, corresponding to two perpendicular bounces on the two almost-flat arcs . This region is bounded by a curve $c_0(b)$ and the value $c_0(b)=\frac{2}{1-b}$ can be easily calculated, see Figure~\ref{f:param}.

\begin{figure}[tp]
\centering
\includegraphics[scale=0.4]{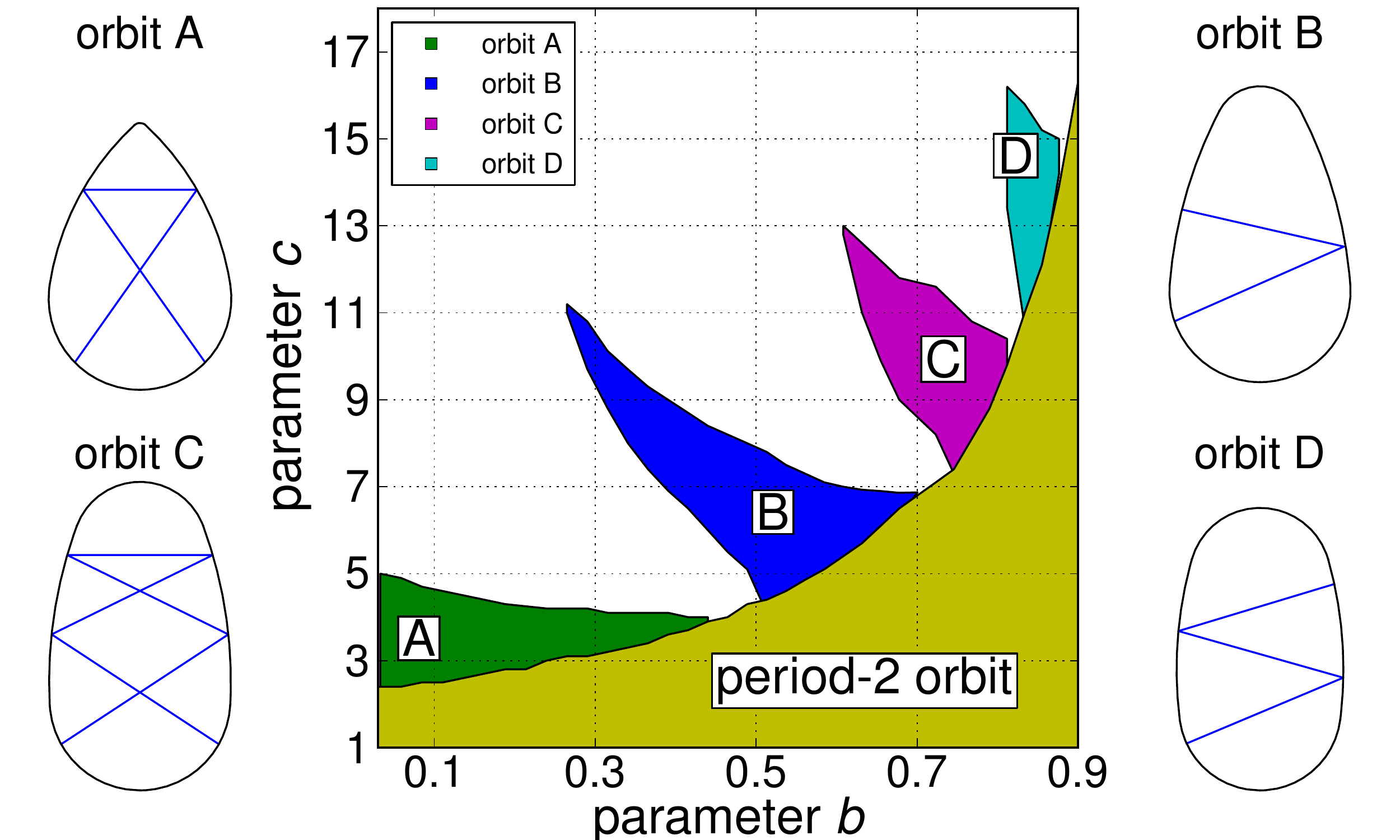}
% \caption{Stability regions of islands around certain periodic orbits. }
\caption{Regions of stability of certain types of periodic orbit. Each shaded region depicts the numerically-determined region of
parameter space in which the corresponding labelled orbit type (with the same topology, or coding) is stable.}
\label{f:param}
\end{figure}

For $c > c_0(b)$, the period-2 orbit and its corresponding island disappear, but other islands, corresponding to periodic orbits of higher periods,
%and various geometrical shapes
appear in certain regions of the $(b,c)$ plane. These islands generally appear in a certain window of values of $c$ for a given, fixed $b$. Such stability regions for certain orbits of low period are shown in figure~\ref{f:param}. For example, in the (blue) B region in the parameter space of figure~\ref{f:param}, there exists an island around the period-4 orbit shown at the top right, for which the bounces on the bottom and left arcs are perpendicular (see also figure~\ref{f:orbits}(a)).

\subsection{Geometric destruction and period doubling of periodic orbits}
\label{s:period_doubling_geom_destruction}

We proceed to investigate some features of the phase portrait, in particular, the regions of existence and stability for the periodic orbits. Bifurcation phenomena in piecewise-smooth systems are the subject of intensive studies -- see e.g.\ \cite{di2008piecewise, simpson2009simultaneous} and references therein. It is beyond the scope of this paper, and in any case not our main goal here, to provide a detailed study of the bifurcations occurring in the system; instead, we give a short description of the main features we have observed (see \cite{dullin_two-parameter_1996} for a similar discussion).

Based on our observations, there are two kinds of dynamical phenomena that mainly determine whether an island of stability is formed around a periodic orbit. On the one hand, the singularities of the boundary play an important role, since they influence the very existence of periodic orbits.
% In this section, we the phenomena which determine the regions of existence and stability of periodic orbits; see also \cite{dullin_two-parameter_1996} for related discussions.
To describe these orbits succinctly, we introduce a coding of four letters
$\{b, r, t,  l\}$, denoting bounces on the bottom, right, top and left arcs, respectively, and describe periodic orbits according to a finite code of length equal to the period of the orbit. Even though this code is not unique, it is useful since it highlights the role of the singularities, as far as the existence
and stability features of periodic orbits are concerned.
%\begin{incom}
% This code is non-unique, e.g. for period-2 diametrical orbits if the bottom arc is more than a semi-circle. In fact, it is not clear to me that it is really %worth introducing this code.
%\end{incom}

For example, consider again the period-$2$ orbit with code $(r,l)$, which consists of consecutive perpendicular bounces on the two
almost-flat arcs, and which exists for $c < c_0(b)$. As $c$ tends to $c_0(b)$ from below, the locations of the bounces on the sides move closer and closer to the ends of the almost-flat arcs, and finally reach the lower corner points (singularities) when $c=c_0(b)$. For larger values of $c$, this period-$2$ orbit with code $(r,l)$ thus ceases to exist, since there are no points on the almost flat arcs with mutually parallel tangent lines. We   refer to this as \emph{geometric destruction} of the periodic orbit.

\begin{figure}[tp]
  \centering
    \subfigure[Period-$4$ orbit with $c = 8.59$.]
      {\includegraphics*[scale=0.4]{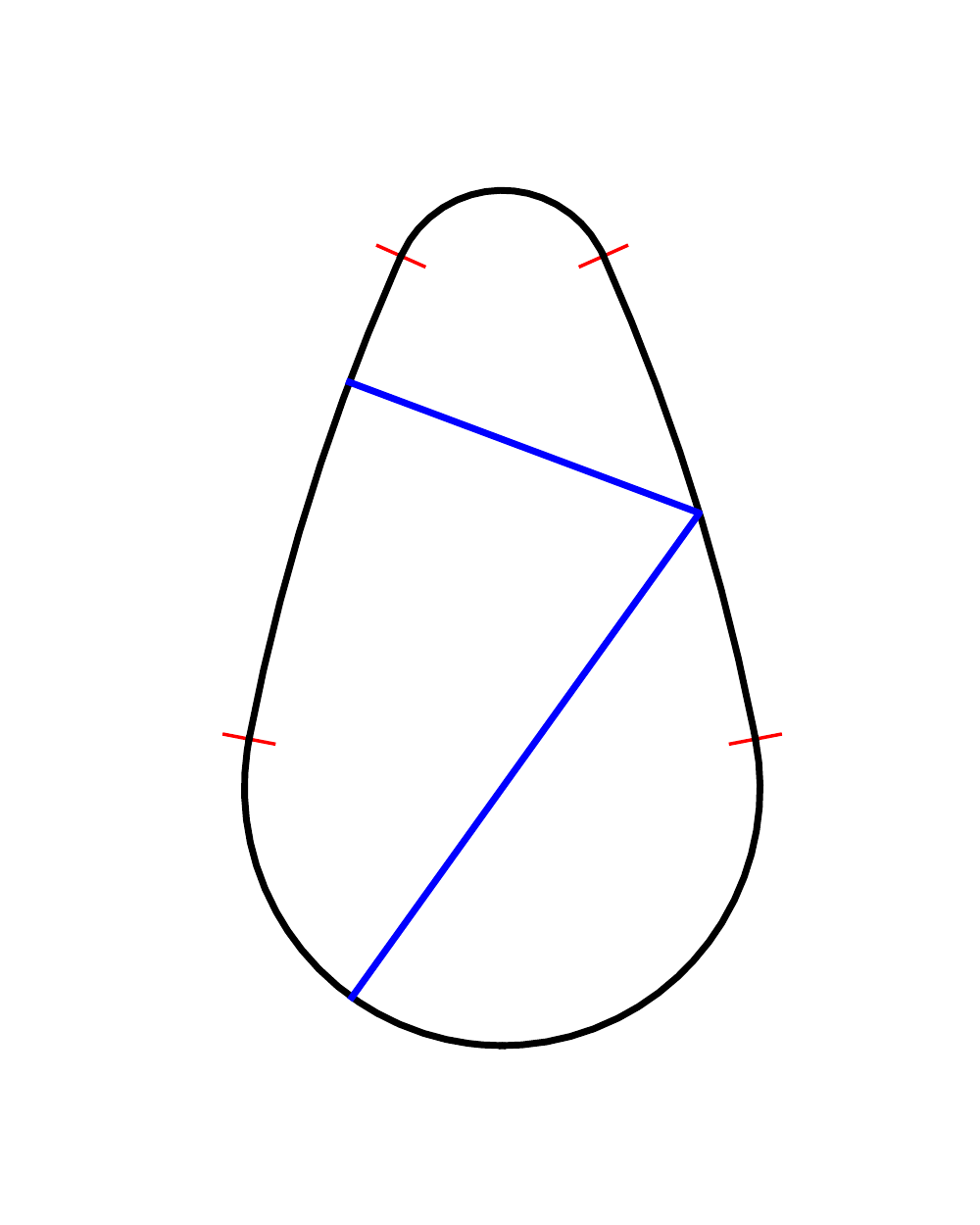} }
        \qquad
    \subfigure[Period-$8$ orbit with $c = 8.61$.]
      {\includegraphics*[scale=0.4]{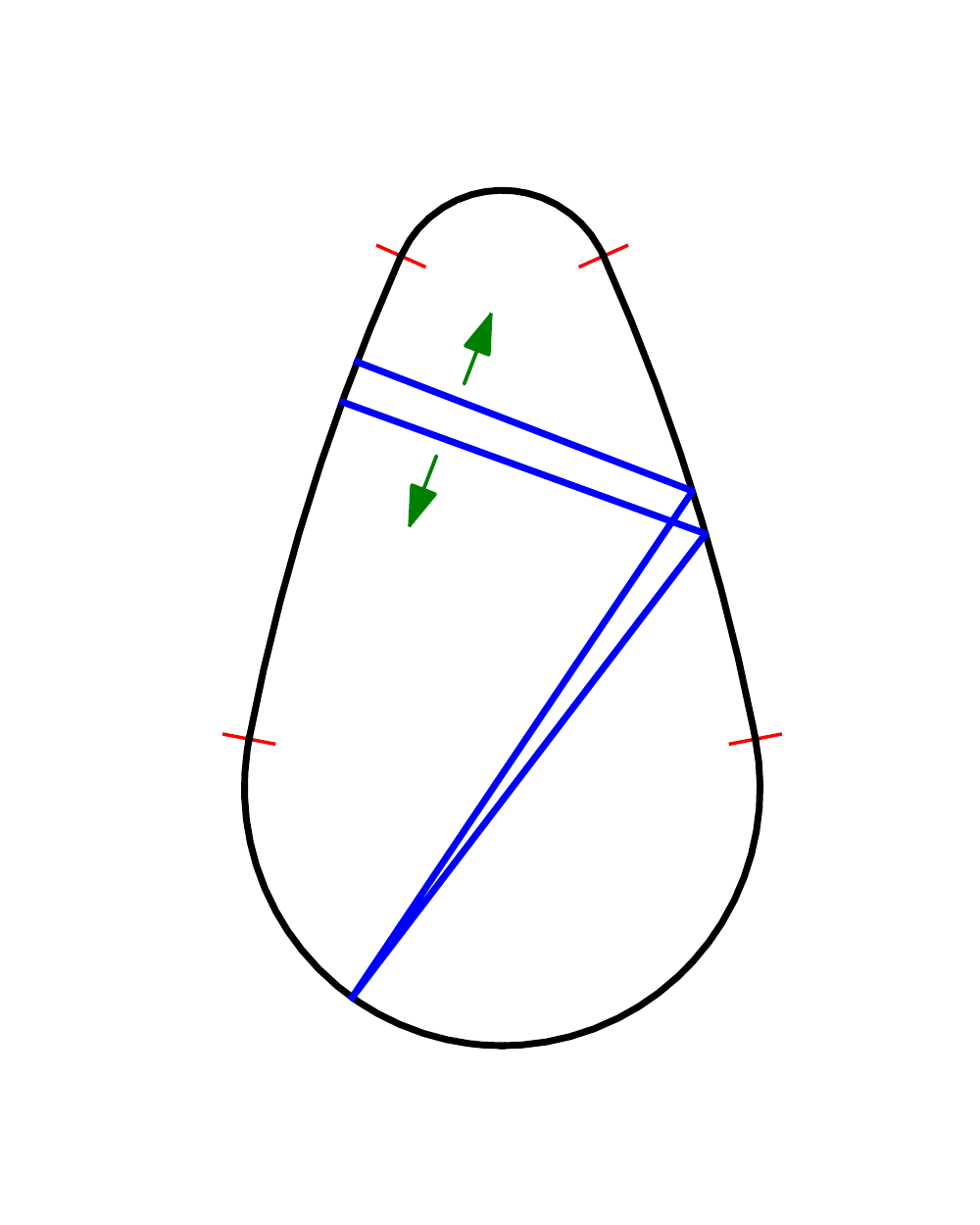} }

  \caption{Period-doubling bifurcation of a period-4 orbit for $b=0.4$ as $c$ is varied. The (red) thin straight lines indicate the singularities in the boundary of the squash, where the circular arcs are joined. The arrows indicate the direction in which the trajectory moves as $c$ increases. }
  \label{f:orbits}
\end{figure}

However, geometric destruction is not the only mechanism by which an island corresponding to a periodic orbit may disappear. For example, figure~\ref{f:orbits}(a) shows a period-$4$ orbit, around which an island of stability exists in the range
$6.7 \lesssim c \lesssim 8.7$ for for $b=0.4$. When $c \simeq 8.7$, the orbit loses stability, even though its points of collision are still located far from the singularities. In this case, the disappearance of the island is rather related to the stability properties of the periodic orbit. KAM theory can only account for islands near \emph{elliptic} periodic points, i.e.\ if the stability parameter (see equation~(\ref{f:linstab}) and the following paragraph) satisfies $|s|<2$.

% It is interesting to ask what happens in the vicinity of a periodic orbit when it loses stability ($|s|$ increases above $2$).
In  the \emph{smooth} setting, it is known that when a periodic orbit loses stability ($|s|$ increases above $2$)  \emph{period-doubling bifurcations} may be observed; see eg.\ \cite{MacK}. As observed in
\cite{dullin_two-parameter_1996}, this can also occur in billiard-type dynamics, provided the periodic orbit stays away from the singularities.

% \begin{incom}
% It would be more appropriate to cite something more recent than \cite{MacK} but I am not sure what to suggest. Do you have an idea?
%
% \end{incom}

\begin{figure}[tp]
  \centering
    \includegraphics[scale=0.42]{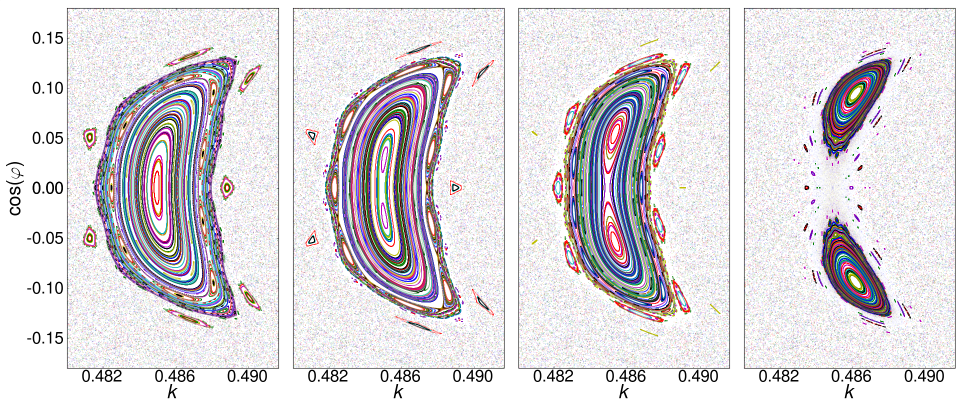}
    \caption{Phase space structures for the period-doubling bifurcation of the elliptic periodic orbit shown in figure~\ref{f:orbits}(a), highlighting the splitting of the corresponding island. Parameter values $b=0.4$ and $c$ = 8.59, 8.61, 8.63 and  8.69 from left to right.}
    \label{f:doubling}
\end{figure}

\begin{figure}[tp]
  \centering
      \includegraphics[scale=0.6]{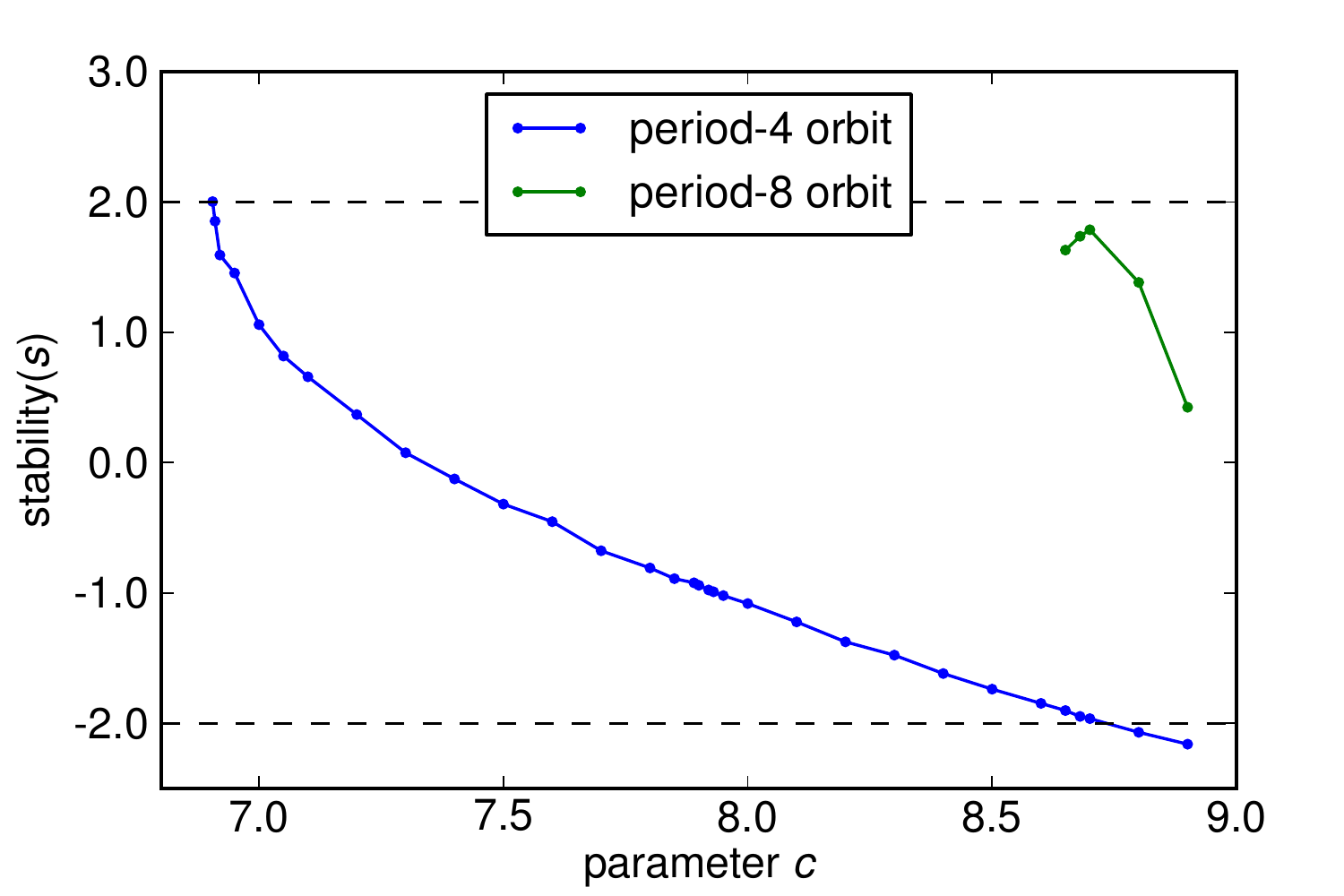}
  \caption{Dependence of the stability on $c$ for the period-4 and period-8 orbits shown in figure~\ref{f:orbits}.}
  \label{f:stab}
\end{figure}

Consider again the period-4 orbit studied above for $b=0.4$. As $c$ increases from $c \simeq 6.7$ to $c \simeq 8.7$, we find that $s$ decreases from $2$ to $-2$; see figure~\ref{f:stab}. When $c$ reaches the critical value $8.7$, the shape of the island changes and it splits into two components, the centers of which are consecutive points of a period-$8$ orbit, shown in figure~\ref{f:orbits}(b).
The changes in shape of the corresponding islands which surround the periodic orbits is shown in figure~\ref{f:doubling}.

 The period-$8$ orbit is elliptic even for larger $c$, when the original period-4 orbit is already hyperbolic, so the island is formed around this elliptic orbit of double period for these parameter values.
%  Figure~\ref{f:stab} shows how the stability properties of the period-4 orbit and the period-8 orbit change when $c$ is increased.
Similar period-doubling bifurcations occur for other periodic orbits whose disappearance is  not due to geometric destruction.
%Correspondingly, the phase portrait of the island also changes, as shown in Figure~\ref{f:doubling}.

As $c$ is increased even further, the geometry of the period-$8$ orbit is deformed, and soon the location, for at least one of the bounces, reaches a corner,  resulting in geometric destruction of the island.

Summarizing, we can see that islands are subject to the combination of two effects: geometric destruction caused by the singularities, combined with the generic features characteristic of smooth systems with mixed phase space. Indeed, the presence of the
singularities modifies the picture significantly. See also \cite{dullin_two-parameter_1996, AMS} on similar phenomena in the billiard setting.

% \subsection{Larger values of $c$}

\subsection{Main conjecture: a class of ergodic convex billiards}
  \label{s:main_conjecture}
%
% IS THE FOLLOWING SUMMARY NEEDED? The observed phenomena can be summarize as follows:
%
% \begin{itemize}
%     \item For not too small $c$, ergodic components of positive Lebesgue measure seem to fill most of the phase space. In the complement of this ergodic components  we observe the typical mixed phase space characteristic of KAM theory: elliptic islands and chaotic orbits coexist in a topologically complicated way. For $c\ge 1.5$, we observe a single, dominant ergodic component.
%
%     \item For a given value of $b$, if we increase $c$, then the islands shrink in size and/or disappear due to geometric destruction and also they can split themselves due to period doubling bifurcations; this islands with higher period are more sensitive to geometric destruction. For high enough value of $c$, it appears that no island survives.
%
% \end{itemize}

In this section we concentrate on the case of large $c$ for some $b<1$ fixed, and formulate our main conjecture concerning this region of  parameter space.

The tendency is that as $c$ increases, the islands of stability are formulated around orbits of higher period.
% \begin{incom}
% IS THIS TRUE? Yes I think it is. A heuristic reasoning (which I do not think we should include in the paper): as $b<1$, no periodic orbit can close on the almost flat arcs, it must bounce at least on the bottom arc as well. Hence there is an orbit with the same geometry for $c=\infty$ as well, in which case it would be hyperbolic. For $c<\infty$ such an orbit can only be elliptic if the tangent map along it is sufficiently far away from the hyperbolic tangent map of the $c=\infty$ case. TO achieve this for big $c$, we need a high number of bounces on the almost flat arcs so that the cumulative effect takes the tangent map of the orbit from the hyperbolic to the elliptic side.
% \end{incom}
% OK, that seems convincing to me
Typically, orbits of higher period are more sensitive to geometric destruction, since the presence of more bounces leads to more possibilities for one of the collision points to collide with a singularity when the parameters are changed.  Thus the area in phase space of higher-period islands tends to decrease with increasing period.
For this reason, the region in the parameter space where period-doubling bifurcation phenomena can actually be observed is quite limited.

Furthermore, as the parameter $c$ is increased even further, it appears that \emph{all} islands  disappear, and that the system thus becomes ergodic.
These numerical observations motivate the following conjecture:

\begin{conjecture}
  \label{conj}
    For any $b<1$, there exists $\hat{c}(b)$ such that for any pair of parameters $(b,c)$ with $c>\hat{c}(b)$ the billiard dynamics are ergodic.
\end{conjecture}
That is, once the generalised squash billiards are close enough to skewed stadia, we conjecture that they are ergodic.
%
% \begin{remark}
%   It is important to point out, however, that the period-2 orbit between the almost-flat arcs is \emph{not} necessarily responsible for the ``last disappearing island'' for increasing $c$. In particular, if $b<0.95$, for certain values of $c$ for which the bottom arc is larger than a semi-circle, we have observed phase portraits which suggest that the table is not yet ergodic, since islands lying around other elliptic periodic orbits with higher period are present. Nonetheless these islands also disappear if $c$ is further increased. See also section~\ref{s:param_space} on these phenomena.
% \end{remark}
There is an obvious lower bound $\hat{c}(b) \ge c_0(b)$ on the  region of ergodicity in parameter space;
both quantities tend to $\infty$ as $b$ tends to $1$.

Similar conjectures have been made for certain regions of parameter space in other billiards with piecewise-smooth boundaries \cite{dullin_two-parameter_1996}.
Here, however, we provide heuristic arguments, and confirmation using a more powerful numerical method, in the next section.

\section{Heuristic and numerical evidence for Conjecture~\ref{conj}}
 \label{sec:evidence-supporting-conjecture}
In this section, we support Conjecture~\ref{conj} with heuristic and further numerical arguments. Even though a rigorous proof, which appears to be a technically challenging task, is currently not available, we believe that we capture the key phenomena. We also apply the powerful numerical method of Lyapunov-weighted dynamics to perform a more stringent search for elliptic islands.

\subsection{Heuristic arguments}\label{s:heur}

We compare the dynamics of the model for large $c$ with the limiting case $c=\infty$, i.e.\ the skewed stadia, which has been studied extensively in the literature, for any value of $b$, and is known to be ergodic; see, for example, \cite{CZ2}.
%  In this section we provide some heuristic reasoning supporting Conjecture~\ref{conj}.

\paragraph{Comparison with the limiting case $c=\infty$.}
Consider the subset
$\bM\subset M$ consisting of phase points $x=(k,\phi)$ for which (i) $k$ belongs to one of the circular arcs (not to the almost-flat ones), and (ii) the $\phi$ coordinate ensures that for $(k',\phi')=T(k,\phi)$, the point $k'$ belongs to an arc different
from the one on which $k$ is located. It is proved in \cite{CZ2} that for $c=\infty$,  the first return map to
$\bM$ is \emph{uniformly hyperbolic}.

Now, for fixed $b$ the billiard tables corresponding to finite $\bar{c}=c\gg 1$ and to $c=\infty$ are (piecewise) $C^2$-close.
Let us denote the corresponding billiard maps by $\Tc$ and $\Ti$, and the return maps to $\bM$ by $\Tvc$ and $\Tvi$, respectively. Now we recall some properties of the billiard map from \cite{CM}; see also formula (\ref{f:linstab}) in the Appendix.
$C^2$-closeness of the billiard tables would imply that the maps $\Tc$ and $\Ti$ are $C^1$-close; however, the billiard map may have unbounded derivatives corresponding to tangential collisions.
Thus, $\Tc$ and $\Ti$ are $C^1$-close unless $\phi \simeq 0$ or $\phi \simeq \pi$.

We would like to conclude that (apart from tangencies) $\Tvc$ is also $C^1$-close to $\Tvi$, and thus, $\Tvc$ is also uniformly hyperbolic (as uniform hyperbolicity is a $C^1$-open property). However, closeness of the $T$'s implies closeness of the $\bT$'s  only if phase points $x\in \bM$ are considered for which $\bT x=T^{n(x)}x$ such that $n(x)$, the number of iterates needed to return to $\bM$, is \emph{uniformly bounded}. Thus, we need to consider the complement of this set: points with unbounded return time, which are exactly the quasi-integrable trajectories. At this point the differences between the $b<1$ and $b=1$ cases play an important role, as follows.
\begin{enumerate}
\item For \emph{bouncing points} $x = (k,\phi)\in \bM$, returns to $\bM$ consist of a high number of consecutive collisions on the almost flat arcs. If $b=1$, then such orbits may spend an unbounded number of iterations bouncing close-to-perpendicularly on the almost-flat arcs. However, if $b<1$, then the number of such bounces, and thus the time needed to return to $\bM$, is uniformly bounded (the bound, of course, depends on the actual value of $b$, which affects the value of $\hat{c}(b)$ in Conjecture~\ref{conj}).

\item \emph{Diametrical quasi-integrable} motion, when the trajectory moves back and forth between two opposite points of a circle (see the end of section~\ref{sec:mod}), may also correspond to unbounded return time. This phenomenon, however, is restricted to the bottom arc of the $b<1$ case, which has the same geometry for $\Tc$ and $\Ti$. More precisely, the almost-flat arcs play a negligible role: returns to $\bM$ consist of a diametrical trajectory segment (that can be arbitrary long) and a single bounce on one of the almost flat arcs (cf. \cite{CZ2}). This allows us to compare directly the derivatives of the return maps $\Tvi$ and $\Tvc$ in such points. Throughout, we use the 
    local orthogonal section to describe tangent maps (cf. \ref{s:numapp}). Consider $x\in\bM$ with return time $n(x)=N+1$ performing diametrical quasi-integrable motion, and let us denote the derivatives of the return maps of the $c=\infty$ and $c=\bar{c}$ cases by $D_N=D_x(\Tvi)$ and $\bar{D}_N=D_x(\Tvc)$, respectively. Then $D_N=B_N \cdot A$ and $\bar{D}_N=B_N\cdot \bar{A}$, where the matrix $B_N$ is the tangent map corresponding to $N$ consecutive diametrical bounces on the bottom arc, while the matrices $A$ and $\bar{A}$ are the tangent maps corresponding to the single bounce on the flat arc of the $c=\infty$ case and on the almost flat arc of the $c=\bar{c}$ case, respectively. Then we have:
    \begin{itemize}
    \item $D_N$ is strongly hyperbolic, expanding unstable vectors by at least a factor $k_1\cdot N$ for some numerical constant $k_1>0$ (see \cite{CZ2}).
    \item All elements of the matrix $B_N$ have absolute value less than $k_2\cdot N$ for some numerical constant $k_2>0$. (This can easily be checked by direct computation based on equation~(\ref{f:linstab}).) 
    \item Choosing $\bar c$ big enough, the difference of the matrices $A$ and $\bar{A}$ can be made arbitrarily small: that is, for any $\eps>0$ there exist some finite $c'$  such that whenever $\bar{c}>c'$ we have $||A-\bar{A}||<\eps$ (which again follows from direct computation). 
    \end{itemize}
    The above three facts imply that, choosing $\bar{c}$ big enough, $\bar{D}_N$ expands unstable vectors  at least by a factor $k_3\cdot N$ for some positive constant $k_3$. In particular, at points that give rise to diametrical motion, uniform hyperbolicity of the return map persists for finite, large enough $\bar{c}$.
\item Thus we only need to show that the third type of quasi-integrable motion,  \emph{sliding trajectories}, that is, points with $\phi \simeq 0$ or $\phi \simeq \pi$, does not spoil the picture.
\end{enumerate}

For brevity we introduce the coordinate $\psi$ where
$\psi=\phi$ if $\phi \simeq 0$, and $\psi=\pi-\phi$ if $\phi \simeq \pi$. The arguments
above can be summarized as follows: fix the parameters $(b,c)$ where $b<1$ and $c>\hat{c}(b)$.
Then there exists $\psi_{b,c}$ such that $\bT$ is uniformly hyperbolic for any $x\in\bM$ for which
$\psi>\psi_{b,c}$. Furthermore, for any fixed $b<1$, $\psi_{b,c}\to 0$ as $c\to \infty$.

% \medskip\noindent
\paragraph{Analysis of sliding trajectories.} To complete our heuristic argument,
we show that for $c$ large enough, trajectories
leave the sliding region of the phase space (the region where $\psi \simeq 0$)
within uniformly bounded time.

Note that $\psi$ remains constant as long as the sliding trajectory
collides with the same arc. Thus what we need to understand is how the value
of $\psi$ changes when the trajectory bypasses a corner point (and thus switches
to a neighboring arc). There are two types of transitions to be considered:
switching from the bottom (or the top) arc to an almost-flat arc
(an $R\to \Ri$ transition), or from an almost flat arc to the bottom (or the top)
arc (an $\Ri\to R$ transition).

Consider such a transition and denote the value of the collision angle before and after the corner point by $\psi$ and
$\psi'$, respectively. To describe configurations, it is more convenient
to introduce one more quantity, $\beta$, the angular distance of the last bounce on the first arc from the corner point (the distance in
terms of the coordinate $k$ divided by the radius of the circular arc). We have
$0\le \beta\le 2\psi$. First let us consider an $\Ri\to R$ transition.
Using elementary geometry -- sine theorems for the two triangles which are
delimited by the radial segments connecting the center of the arc with the location of the last (first) bounce,
the radial segment pointing to the corner point, and the trajectory
% , see Figure~\ref{f:surran}
-- we find
$$
\frac{1}{c}\cos\psi'+\left(1-\frac{1}{c}\right)\cos(\beta-\psi)=\cos\psi.
$$

Now, as $\psi$ (and thus $\beta$) is small, we approximate the cosines by
second-order Taylor polynomials, and as $c$ is large, we neglect corrections
proportional to $\frac{1}{c}$, to obtain
$$
\psi'^2\approx\psi^2+c\beta(2\psi-\beta).
$$
This means that $\psi'\sim c\psi$ at $\Ri\to R$ transitions,
\emph{with high probability} -- unless the last bounce on the almost-flat arc, or the
first bounce on the bottom arc is extremely close to the corner point.
%
% \begin{figure}[tp]
% \centering
% \includegraphics[width=8cm,clip]{surran.pdf}
% \caption{Analysis of sliding trajectories}
% \label{f:surran}
% \end{figure}

On the other hand, for $R\to \Ri$ transitions a completely analogous argument yields
$$
\psi'^2\approx \psi^2-\beta(2\psi-\beta),
$$
which means that $\psi'\sim \psi$ with high probability in this case.

As $R\to \Ri$ and $\Ri\to R$ transitions alternate, for a typical sliding trajectory
the value of $\psi$ exhibits a big kick at every second transition, and changes
moderately at the intermediate occasions. This results in escaping from the sliding region
after a couple of transitions.

These phenomena have  been tested by convincing simulations,
which can be reproduced by the reader using the program available at \cite{program}.
Note that on the phase portraits produced by this program,  $k$ is displayed as the configurational
(horizontal) coordinate instead of $\beta$ -- this should be kept in mind when comparing the simulated
trajectories with the above calculations.

\subsection{Lyapunov-weighted dynamics \label{sec:lyap}}

Further support for Conjecture~\ref{conj} is provided by numerically searching the two-dimensional parameter space for pairs $(b,c)$ where no elliptic islands are found, since the system exhibits a dichotomy: the system is (presumably) ergodic if there are \emph{no} elliptic islands,  whereas the \emph{presence} of elliptic islands -- even small ones -- implies that the system is not ergodic.

The simplest method to search for elliptic islands consists of sampling from a grid of initial conditions in phase space and estimating the Lyapunov exponent or an equivalent indicator \cite{lyapunov_indicator} for each one. However, this is unreliable and time-consuming, since the number of initial conditions needed to locate an island is inversely proportional to the area of the island in phase space.

%
% To obtain a stronger numerical test of ergodicity, it is necessary to use a more reliable method of detecting elliptic islands, since we have the dichotomy that the presence of elliptic islands ruins ergodicity, whereas the absence of elliptic islands (presumably) implies it.
% Sampling over random initial conditions is unreliable and time-consuming, since the number of initial conditions needed to locate one grows inversely proportional to the size of the island.
%
% As previously remarked, atypical trajectories are important because they reveal essential information about the dynamics. In this section we locate these trajectories numerically for different parameter values of the squash billiard. The idea of this is to obtain numerical evidence that support the ergodicity of the system for c greater than some critical value $\hat{c}(b)$, as stated in conjecture \ref{conj}. That is, we are going to look for regular regions ($\alpha=-1$) and we would expect to find no elliptic islands, otherwise the system would not be ergodic.

A more powerful approach to locate  regions of phase space with regular (non-chaotic) dynamics is the  \emph{Lyapunov-weighted dynamics}  method (LWD), introduced in ref.~\cite{lwd_articulo}.  The idea of this method is to evolve a population of walkers under a modified version of the dynamics, chosen so that the cloud of walkers spontaneously \emph{concentrates} in regions of phase space which have a large (respectively small) Lyapunov exponent, according to whether a parameter $\alpha$ of the method is positive (respectively negative) \cite{lwd_articulo}.

To do so, each walker follows the underlying deterministic dynamics of the system under study, but perturbed by a small random noise of strength $\epsilon$.
Each walker also carries a tangent vector, which evolves under the tangent dynamics of the map. The local stretching factor of this tangent vector is calculated, as in the standard calculation of Lyapunov exponents, and walkers are killed or copied (``cloned'') at a rate which is
proportional to ($\alpha$ times) the stretching of their tangent vector \cite{lwd_articulo}. This process can be shown to lead to the desired effect of 
the cloud of walkers ``highlighting'' the regions of chaotic or regular behaviour, depending on the value of $\alpha$ chosen \cite{lwd_articulo}.
% have the desired effect, by which the walkers are led to more or less stable regions of phase space \cite{lwd_articulo}.
% After a given time interval, walkers that instantaneously do not have the desired behavior are killed and those with the desire behavior are copied with a rate proportional to ($\alpha$ times) the stretching of their tangent vector.
Details of the application of the LWD method to billiard models will be discussed in a forthcoming publication \cite{JHT-DPS}.

% The overall effect of this modified dynamics is that the
% A positive (negative) value of $\alpha$ tends to favor orbits with a large (small) Lyapunov exponent \cite{lwd_articulo}.
% % Thus the population of walkers has a tendency to ``highlight'' the regions of desired chaotic or regular behaviour, depending on the value of $\alpha$ chosen.
% An alternative method to achieve this was proposed in ref.~\cite{Magnasco}.

\begin{figure}[tp]

  \centering
    \subfigure[$\alpha=-1$]
    {
      \includegraphics[scale=0.31]{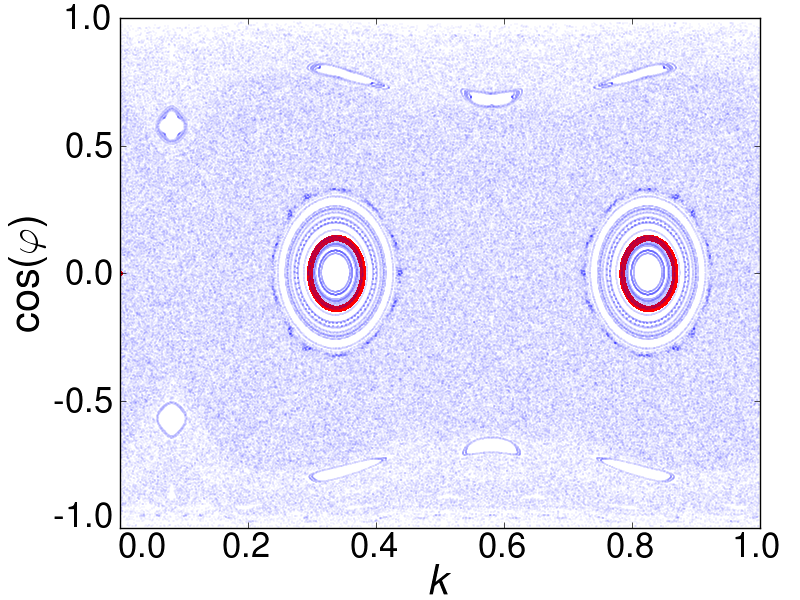}
      \label{fig:mixed_islands}
    }
    \subfigure[$\alpha=+1$]
    {
      \includegraphics[scale=0.31]{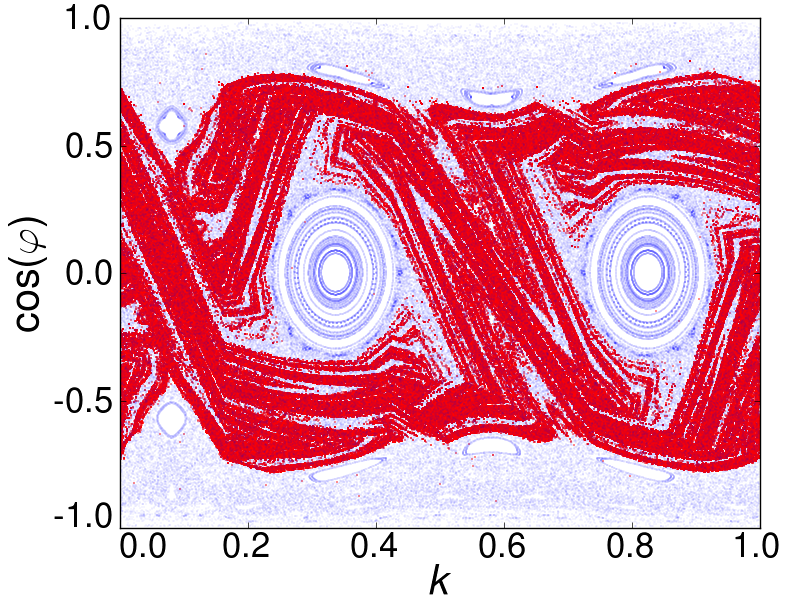}
      \label{fig:mixed_ergodic}
    }
   \caption{Results of applying LWD to a squash billiard with parameters $b = 0.6$ and $c= 1.6$, searching for both (a) regular and (b) chaotic regions. The thin (blue) points represent the evolution of randomly chosen initial conditions that evolve with the normal dynamics of the squash billiard, thus giving the standard representation of the phase space of the system. The (red) darker points show the superposition of the phase space locations during the last 100 collisions of 1000 walkers evolved under LWD for 10000 collisions. }
   \label{fig:mixed}
\end{figure}

Figure \ref{fig:mixed} illustrates the results of applying the LWD method to a generalised squash billiard, searching for both regular ($\alpha=-1$) and chaotic ($\alpha=+1$) dynamics.
When $\alpha=-1$, the walkers concentrate in one of the elliptic islands of the phase space, whereas when $\alpha$=1, the walkers instead stay in part of a chaotic component of the phase space, as shown in figure \ref{fig:mixed_ergodic}. In this case, there is in fact a partial barrier to transport in the phase space, visible around $\cos(\phi) \simeq 0.7$ in figure~\ref{fig:mixed}. The LWD dynamics with $\alpha=+1$
nonetheless only sees part of this chaotic component.

\begin{figure}[tp]
  \centering
      \subfigure[$b = 0.1$, $c= 4$]
	  {
	     \includegraphics[scale=0.31]{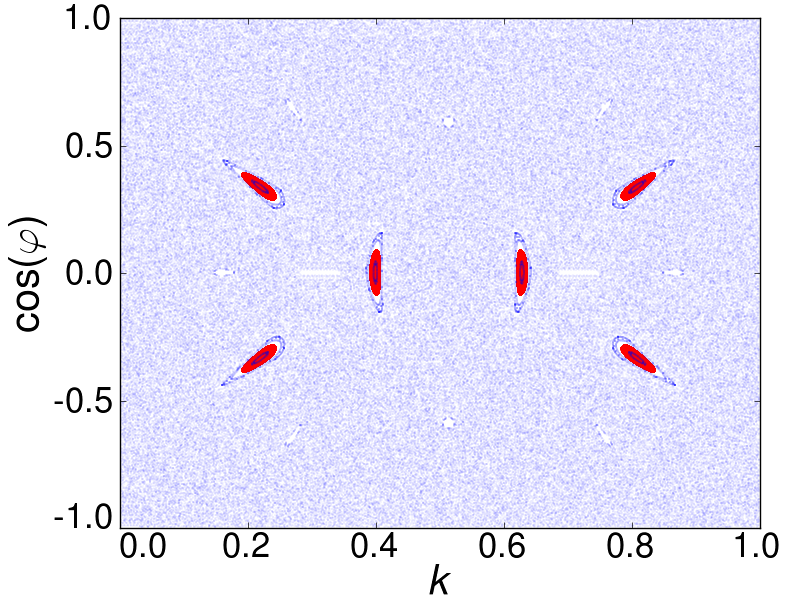}
	      \label{fig:parabolic1}
	  }
      \subfigure[$b = 0.1$, $c= 6$]
	  {
	     \includegraphics[scale=0.31]{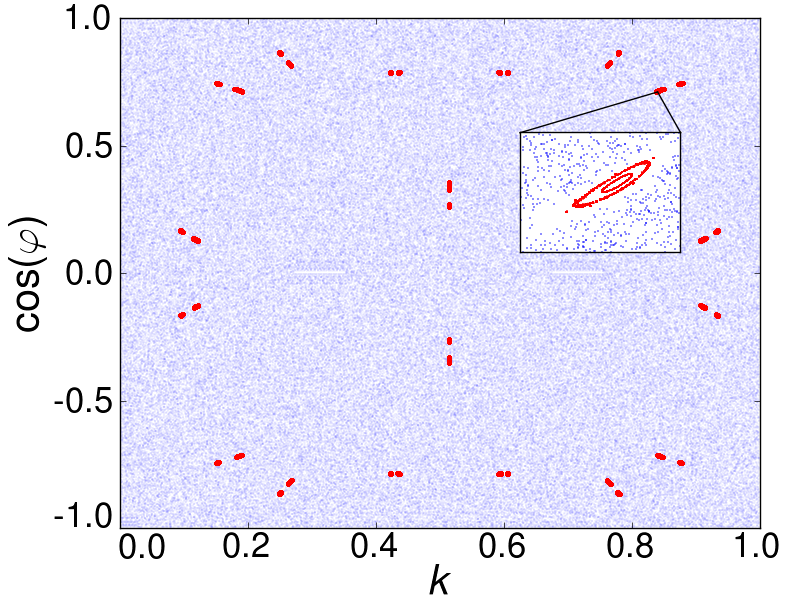}
	      \label{fig:parabolic2}
	  }
  \caption{Results of applying LWD with $\alpha=-1$ to two squash billiards. The superposition of the last 100 collisions of 10000 walkers evolved under LWD for 10000 collisions is shown in (red) darker points on top of the phase space. }
   \label{fig:parabolic_vs_islands}
\end{figure}

% Figure \ref{fig:parabolic1} illustrates the results of applying the LWD method  with $\alpha=-1$  to a generalised squash billiard.

Our approach is thus to exhaustively explore the $(b,c)$ parameter space, looking for elliptic islands by using a modified version of the LWD method \cite{JHT-DPS} with  $\alpha < 0$. As discussed above, in this case the walkers indeed have a tendency to concentrate in the part of phase space which is ``most stable''. Figure~\ref{fig:parabolic1} shows the results of applying LWD to a squash billiard with parameters $b=0.1$ and $c=4$. For these parameter values, there is an elliptic periodic trajectory of period 6, around which there is an elliptic island, but there is also a set of parabolic periodic orbits, corresponding to diametrical bouncing, since the lower circular arc is longer than a semi-circle, as discussed in section~\ref{s:squashes}.%s and showed in figure \ref{f:quasi}.
We see that
%
% This method also has one more advantage: if we are looking for regular regions in phase space ($\alpha=-1$), the elliptic islands are
% preferred over points with parabolic behavior because they are more stable, as can be seen in figure \ref{fig:parabolic_vs_islands},
% where we apply the LWD to a squash billiard with parameters $b=0.1$ and $c=4$. For these parameters, there is a periodic trajectory
% of period six, around which there is an elliptic island; but it also has a set of parabolic periodic points, that correspond to diametrical
% bouncing, since the lower circular arc is longer than a semi-circle, as explained in section  \ref{s:squashes} and showed in figure
% \ref{f:quasi}.
when we apply LWD, the walkers concentrate exclusively around the elliptic island, with none trapped near the parabolic orbits, as shown in figure~\ref{fig:parabolic1}. This preference for the more stable islands is independent of both the size and the period of the island, as indicated in figure~\ref{fig:parabolic2}, where there are two islands of period 18, as well as parabolic bouncing orbits, for $c=6$.

\begin{figure}[htp]
  \centering
      \subfigure[$\alpha=-1$]
	  {
	     \includegraphics[scale=0.31]{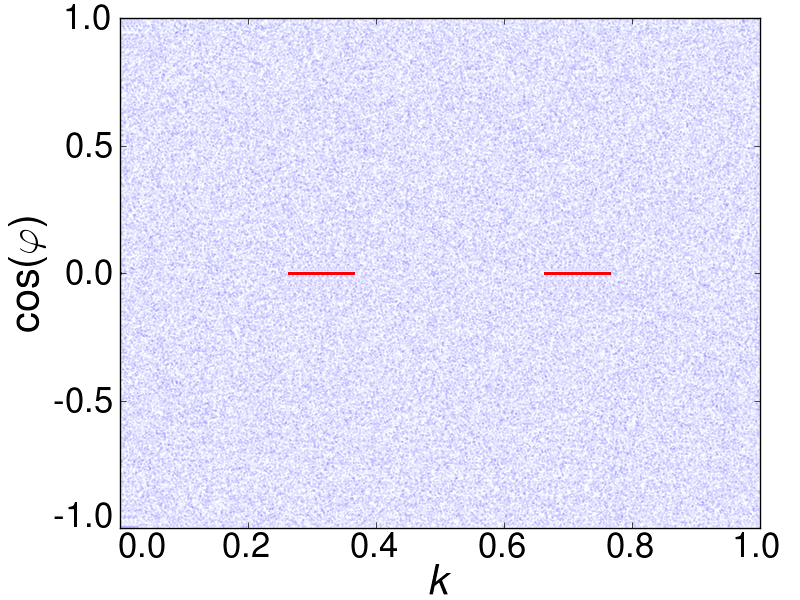}
	      \label{fig:parabolic_and_ergodic1}
	  }
      \subfigure[$\alpha=+1$]
	  {
	     \includegraphics[scale=0.31]{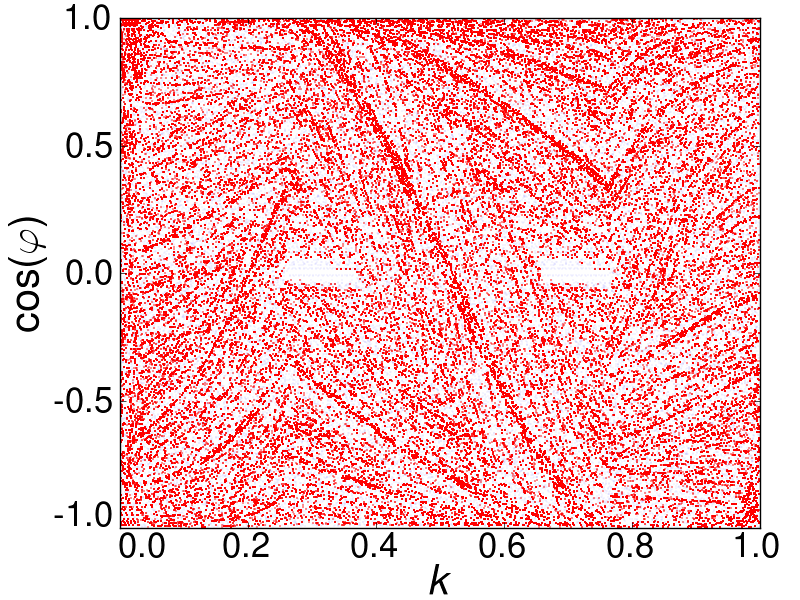}
	      \label{fig:parabolic_and_ergodic1_pesin}
	  }
   \caption{Results of applying LWD to a squash billiard with parameters $b = 0.1$ and $c= 13$. The superposition of the last 100 collisions of 1000 walkers evolved under LWD for 10000 collisions is shown in (red) darker points on top of the phase space.}
  \label{fig:parabolic_and_ergodic}
\end{figure}

If, however, we now increase $c$ further, to $c=13$, then all of the walkers concentrate around the \emph{parabolic} orbits, as shown in  figure~\ref{fig:parabolic_and_ergodic1}. This strongly suggests that there are no longer any elliptic islands in phase space, which would be more stable, and that the system is thus ergodic.

\begin{figure}[tp]
  \centering
      \subfigure[$\alpha=-1$]
	  {
	     \includegraphics[scale=0.31]{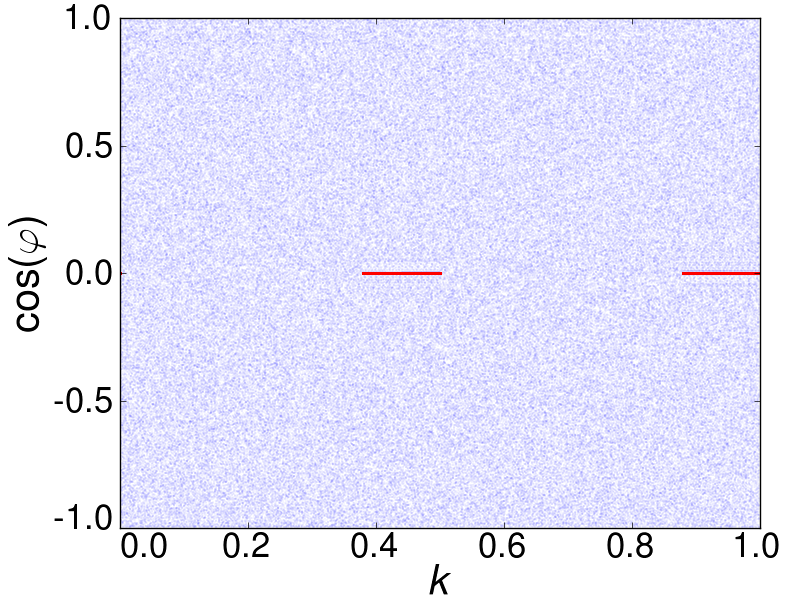}
	      \label{fig:parabolic_and_ergodic_bunim}
	  }
      \subfigure[$\alpha=+1$]
	  {
	     \includegraphics[scale=0.31]{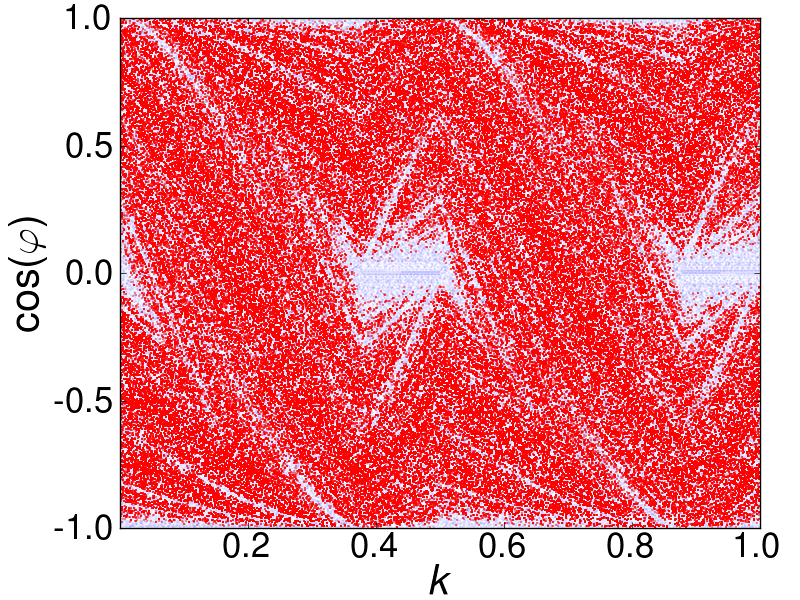}
	      \label{fig:parabolic_and_ergodic_bunim_chaotic}
	  }
\caption{
 Results of applying LWD to an ergodic Bunimovich stadium. The superposition of the last 100 collisions of 1000 walkers evolved under LWD for 10000 collisions is shown in (red) darker points on top of the phase space.}
    \label{fig:parabolic_and_ergodic2}
\end{figure}

Applying LWD with $\alpha > 0$, that is, searching for the chaotic region, instead shows a complicated fractal structure, as shown in figure~\ref{fig:parabolic_and_ergodic1_pesin}. This corresponds to the region of phase space where the rate of expansion is maximised. This region appears to fill phase space in a non-uniform way, avoiding the parabolic orbits.
For comparison, figure~\ref{fig:parabolic_and_ergodic2} shows the results of applying LWD to a system which is known to be ergodic, namely the Bunimovich stadium. Similar behaviour is found as for the squash billiard, which reinforces the idea that the squash is ergodic.

Nonetheless, it is, of course, still possible that  small elliptic islands in phase space still remain in the case of the squash billiard, but whose area is below the threshold which may be detected using the LWD method. This threshold is the consequence of the noisy nature of the method, which we intend to explore in more detail in future work \cite{JHT-DPS}.

%
% \begin{figure}[tp]
%     \centering
%     \includegraphics[scale=0.3]{parabolic_stadium.png}
%     \caption{Results of applying LWD with $\alpha=-1$ to a Bunimovich stadium. The superposition of the last 100 collisions of 1000 walkers evolved under LWD for 10000 collisions is shown in (red) darker points on top of the phase space.}
%     \label{fig:parabolic_and_ergodic2}
% \end{figure}

\begin{figure}[tp]
  \centering
  \includegraphics[scale=0.4]{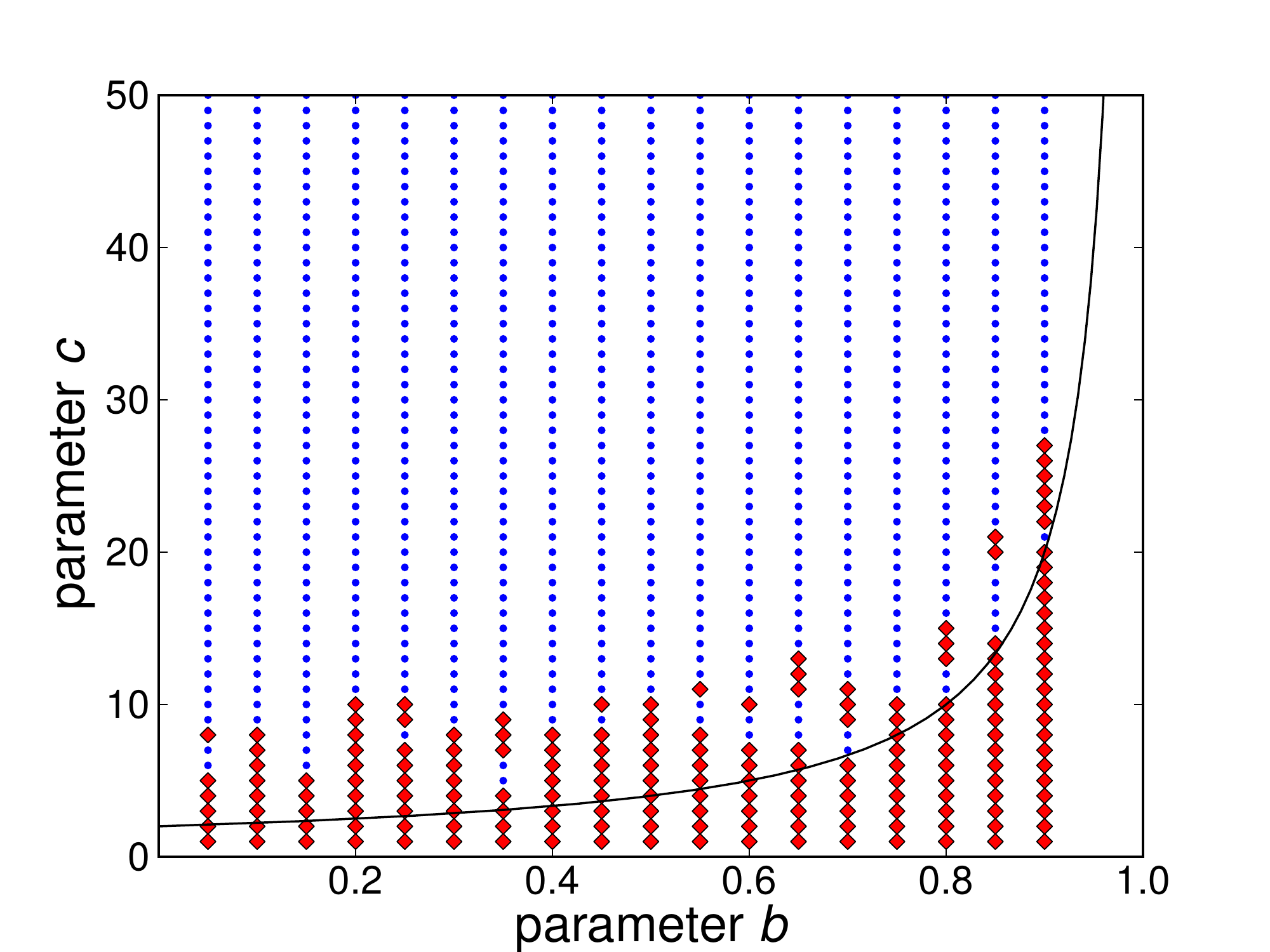}
  \caption{ Parameter space of the generalised squash billiard. LWD with 1000 walkers evolving during 2000 collisions is applied to find regular regions ($\alpha=-1$). Pairs $(b,c)$ that  have elliptic islands found in this way are shown with (red) diamonds; pairs that do not have them are shown with (blue) points. The black continuous curve corresponds to $c_0(b)$.}
  \label{fig:parameter_space}
\end{figure}

% As explained above, by changing the parameters $( b , c )$, the dynamics of the system are changed: the number of ergodic
%  components changes and elliptic islands appear or disappear.
Figure~\ref{fig:parameter_space} shows a sketch of the complete two-dimensional $(b,c)$ parameter space for $c < 50$. Here we
depict the qualitative behaviour found with LWD, i.e.\ whether or not we find elliptic islands for each given pair $(b,c)$. This picture provides strong evidence  for our Conjecture~\ref{conj}, confirming that no islands are found by the method for $c$ above a certain value which depends on $b$.
The disappearance and reappearance of islands observed in the figure as $c$ is increased for certain values of $b$ is presumably related to the interplay of geometric destruction and complex series of period-doubling bifurcations \cite{MacK} (see also section~\ref{s:period_doubling_geom_destruction}).

For even larger values of $c$, we have also taken random values of $c$ and $b$ and we still observe the behaviour shown in figure \ref{fig:parabolic_and_ergodic}, where the walkers concentrate around the parabolic orbits.
As far as the LWD method allows us to exclude the existence of elliptic islands, we thus obtain further evidence
% , although far from conclusive,
pointing towards our Conjecture~\ref{conj} that  the system is ergodic for $c$ greater than some critical value $\hat{c}(b)$.
% Although this is not conclusive, it constitutes numerical evidence that the system becomes ergodic for high enough values of c, and this supports conjecture \ref{conj}.

\section{Conclusions}
\label{sec:conclusions}
We have studied the dynamics of a two-parameter class of generalised squash billiards, which interpolates between completely
integrable and completely chaotic dynamics. We have shown that the dynamical properties in the two-dimensional parameter plane are rather rich, involving a mixture of period-doubling behaviour reminiscent of smooth dynamics, and destruction of orbits caused by collisions with corners of the billiard table.

We have conjectured, based on heuristic arguments which we hope can be made rigorous, and extensive numerical simulations with both standard and Lyapunov-weighted methods, that all elliptic periodic points and their associated islands disappear for tables which are close enough to skewed stadia, thus giving a previously unknown class of ergodic convex billiards. It remains to characterize the parameter space in more detail, and prove the conjecture.

%\subsection{Stability of periodic orbits \label{s:stabper}}

\section*{Acknowledgements}
% DPS thanks Thomas Gilbert for making him aware of the reference \cite{lwd_articulo}.
JHT and DPS acknowledge useful discussions with T.~Gilbert and B.~Krauskopf, and financial support from DGAPA-UNAM PAPIIT grant IN105209.
 PB acknowledges support of the Bolyai Scholarship of the
Hungarian Academy of Sciences, and of OTKA (Hungarian National Found for Scientific Research) grants F60206 and K71693.

\appendix
\section{Construction of generalised squash billiards}

\begin{figure}
  \centering
      \includegraphics[scale=0.35]{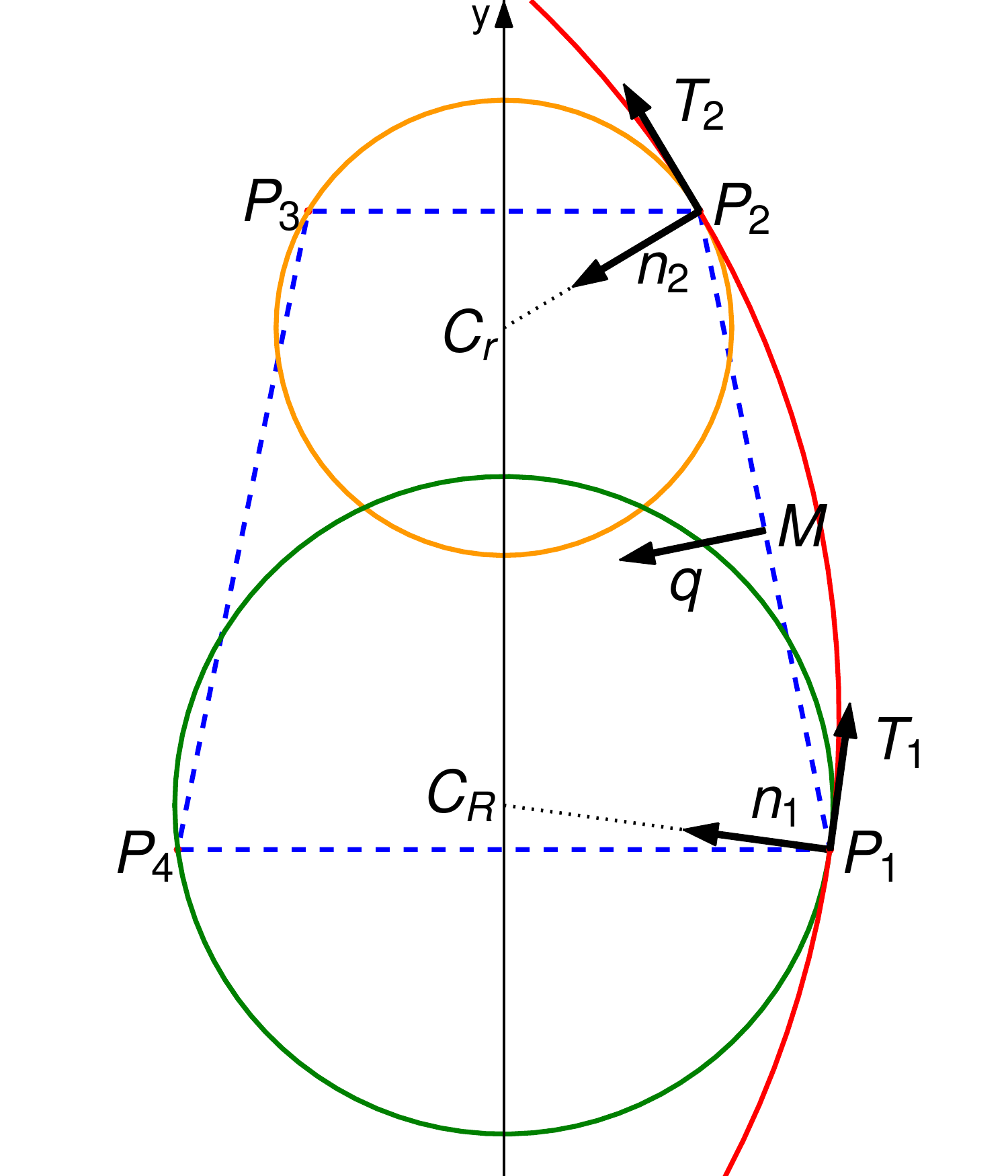}
      \caption{Geometry of generalised squash billiards. The trapezoid is shown in (blue) dashed lines for a generic value of $b$.
The different circular arcs are shown in different grey scales (colours).}
% The circular arc of radius $ r $ is shown in orange, the radius $ R_ \infty $ in red, and the radius $ R $ in green.}
      \label{squash_geom}
\end{figure}

In this appendix, we sketch the geometric construction of generalised squash billiards.
For fixed values of the parameters $ b $ and $ c $, we must determine the radii and centers of the circular arcs that make up the table so that they satisfy the conditions of having common tangents at their points of intersection. To determine these quantities, we use the notation of Figure~\ref{squash_geom}.

Fixing the parameter $ b $ determines the length of the top of the trapezoid and the points $  {P} _1 $, $  {P} _2 $, $  {P} _3 $ and $  {P _4} $.
%We must now determine the radii and centers of the four arcs, which must have common tangents where they intersect.
Since the table is symmetric about the vertical axis, it is only necessary to satisfy the tangent conditions for three arcs, since the right and left arcs have the same radii and their centers are reflected in the vertical axis.

The upper circular arc has radius $ r $ and center $  {C} _r $, the lower arc has radius $ R $ and center $ {C}_R $, and the right arc has radius $ R_ \infty $ and center $ {C}_{R \infty} $. Denote by $ {M}$ the midpoint of the segment joining
$ {P} _1 $ and $ {P}_2 $, and by  $ {q} $ a vector perpendicular to this segment.

By symmetry, $ {C}_r $ is on the $ y$-axis, taking a suitable Cartesian coordinate system. Since $ \| {P}_2 -  {C} _r \| = r $, we have
\begin{equation}
 {C}_r=\left( 0, P_{2y} - \sqrt{r^2-P_{2x}^2} \right) .
\end{equation}
This allows us to calculate the tangent to the upper circular arc $ {T}_2 $ at ${P}_2$, obtaining ${T}_2 = ({C}_r-{P}_2)_\perp$, where $v_\perp$ denotes a vector perpendicular to a given vector $v$.

Let us denote, furthermore, by ${n}_2$ the unit vector perpendicular to $ {T}_2 $.
Since ${C}_{R_{\infty}}={P}_2+s{n_2}$, with $s\in\mathbb{R}$, we have ${T}_2\cdot{C}_{R_{\infty}}={T}_2\cdot{P}_2$.  Moreover, since ${C}_{R_{\infty}}$ is on the line through ${M}$ with direction ${q}$, we find
\begin{equation}
  \label{centro1_squash}
  {C}_{R_{\infty}}={M}+t_2{q}, \qquad \mathrm{where\ } t_2=\frac{{T}_2 \cdot {P}_2 - {T}_2 \cdot {M}} {{T}_2 \cdot {q}} \ .
\end{equation}

With this we can calculate the tangent to the lower arc at the point ${P}_1$, obtaining ${T}_1 = ({C}_{R_{\infty}}-{P}_1)_\perp$.

Finally, ${T}_1 \cdot {C}_R = {T}_1 \cdot {P}_1$ and $C_R$ is on the vertical axis, so
\begin{equation}
  \label{centro2_squash}
  {C}_R = ( 0 , t_1), \qquad \mathrm{where\ } t_1=\frac{{T}_1 \cdot {P_1}}{T_{1y}} .
\end{equation}

These equations give us ${C}_{R}$ and ${C}_{R_{\infty}}$  as a function of  $r$, since ${T}_2$ is a function of $r$ and ${T}_1$ is function of ${C}_{R_{\infty}}$, which is also a function of $r$. Substituting these equations in the definition of the parameter $ c $, we obtain
\begin{equation}
  c=\frac{R_\infty}{R}=\frac{\left \| {C}_{R_{\infty}}-{P}_1 \right \| }{\left \| {C}_{R}-{P}_2 \right \| },
\end{equation}
giving an implicit equation for $ r $ in terms of $ b $ and $ c $:
\begin{equation}
  0=f(r,b,c)=\|{C}_{R_{\infty}}-{P}_1\|-c\|{C}_{R}-{P}_2\| \ .
\end{equation}
This equation may be solved numerically, for example by bisection, to find the value of $ r $ corresponding to given values of $ b $ and $ c $. This can then be substituted back into equations \eqref{centro1_squash} and \eqref{centro2_squash} to get ${C}_{R_{\infty}}$ and ${C}_{R}$, and with this we obtain the radii of the other arcs:
\begin{equation}
  R_\infty=\left \| {C}_{R_{\infty}}-{P}_1 \right \|;
\end{equation}
\begin{equation}
  R=\left \| {C}_{R}-{P}_2 \right \|.
\end{equation}

\section{Numerical methods \label{s:numapp}}
%
% \begin{incom}
%  I am not really sure that we need this section. Maybe we can move it to an appendix?
% \end{incom}

For the two-parameter generalised squashes, to simulate the dynamics
it is sufficient to find the intersection of the line representing the straight trajectory with the different circular arcs, which may be done
by standard methods \cite{GaspardBook}.

A critical role in  the dynamics is played by the periodic orbits,
which reveal much dynamical information. Such orbits may be found by observing the dynamics for several thousand iterates, and searching
for iterates whose coordinates lie in a small neighborhood of the initial condition.
% If the indices of four such consecutive iterates are multiples of the same integer, then the location and length of the periodic orbit can
%  be estimated.
%

 Once a periodic orbit is found, its stability properties can be calculated using the tangent map of the dynamics.
To do so, we use a common convention: instead of the phase space coordinates, the (outgoing) local orthogonal section of the billiard map is used \cite{CM}. The flow coordinates just after collision are taken as the location $r\in Q$ and the velocity $v\in\mathbb{S}^1$; perturbations, up to linear order, are described by the coordinates $(dr,dv)$, where both $dr$ and $dv$ are perpendicular to the velocity $v$.
% To invoke the formulas for the tangent map in the local orthogonal section from \cite{CM},
We fix a phase point $x = (k,\phi) \in M$, and denote its image under the dynamics as $x'=Tx = (k',\phi') \in M$. We denote by $K'$ the curvature of $\Gamma$ at the point $k'$, and by $\tau$ the length of the free flight (the distance between $k$ and $k'$). Then in the
    above-mentioned (outgoing) local orthogonal section coordinates, the image of a tangent vector $(dr,dv)$ at $x$ is the tangent vector $(dr',dv')$ at $x'$, given by \cite{CM}
    \be
    \label{f:linstab}
    \left(
      \begin{array}{c}
        dr' \\
        dv' \\
      \end{array}
    \right)
    = D \left(
      \begin{array}{c}
        dr \\
        dv \\
      \end{array}
    \right)
    =
    \left(
      \begin{array}{cc}
        1 & \tau \\
        \frac{2K'}{\sin\phi'} & 1+ \frac{2K'\tau}{\sin\phi'}\\
      \end{array}
    \right)
    \left(
      \begin{array}{c}
        dr \\
        dv \\
      \end{array}
    \right).
    \ee
The tangent map for a higher iterate, and in particular for the first-return map corresponding to a periodic orbit, can be calculated as the product of several such matrices. Note that in the local orthogonal section coordinates the billiard map is area preserving,
%which is reflected by $D\in SL(2,\mathbb{R})$.
i.e.\ it has determinant $1$. Thus the eigenvalue spectrum of $D$ is characterized by the trace $s$ of $D$: the orbit is hyperbolic, parabolic or
elliptic, if $|s|>2$, $|s|=2$ or $|s|<2$, respectively. %This specifies the algorithm for testing the stability type of a given periodic orbit.
Lyapunov exponents of orbits can be also estimated, as the exponential growth rate of the trace of the matrix $D_n$ giving the tangent  map of the $n$th iterate.
% \end{enumerate}

\section*{References}

\bibliographystyle{alpha}
\bibliography{bib_squash}

\end{document}